\documentclass[preprint,floats,aps, superscriptaddress,nofootinbib,floatfix,showkeys]{revtex4-2}

\usepackage{natbib}
\usepackage{times} 
\usepackage{amssymb,amsmath}
\usepackage{csquotes}
\usepackage{physics}
\usepackage{bm}
\usepackage{nicefrac}
\usepackage{mathtools}
\usepackage{setspace}
\usepackage{booktabs}
\usepackage{upgreek}
\usepackage[per-mode=fraction]{siunitx}

\usepackage[usenames,dvipsnames]{color}
\usepackage[pdftex]{epstopdf}
\usepackage[pdftex]{graphicx}
\usepackage{hyperref}
\hypersetup{pdftitle={Manuscript on } 
    pdfauthor={Kristian Blom and Uwe Thiele}, 
pdfproducer={lateX},
pdfview=FitV,       
pdfstartview=FitB,
linkcolor=blue,     
citecolor=blue,     
urlcolor=blue,      
breaklinks=true,    
colorlinks=true,
citebordercolor=0 0 0,  
filebordercolor=0 0 0,
linkbordercolor=0 0 0,
menubordercolor=0 0 0,
urlbordercolor=0 0 0,
pdfhighlight=/I,
pdfborder=0 0 0,   
bookmarksopen=true,
bookmarksnumbered=true
}
\usepackage{tcolorbox}
\usepackage[dvipsnames]{xcolor}
\DeclareGraphicsExtensions{.jpg, .pdf, .tif, .png}
\usepackage[normalem]{ulem}

\definecolor{reviewchangecolor}{rgb}{1, 0, 0}

\newcommand{\subref}[2]{\hyperref[#1]{\ref{#1}#2}}
\usepackage{amsmath}
\DeclareMathOperator\arctanh{arctanh}

\usepackage{booktabs}
\usepackage{multirow}
\makeatletter
\renewcommand{\p@subsection}{}
\renewcommand{\p@subsubsection}{}
\renewcommand{\@seccntformat}[1]{\csname the#1\endcsname\quad}
\makeatother
\begin{document}
\count\footins = 1000 

\title{Mean-Field Theory of Chiral Active Model B: \\ Arrested Coarsening and Chiral Fingering Instabilities}

\author{Kristian Blom}
\email{\textcolor{blue}{kristian.blom@uni-muenster.de}}
\thanks{ORCID ID: 0000-0001-6582-8448}
\affiliation{Institute of Theoretical Physics, University of M\"unster, Wilhelm-Klemm-Str.\ 9, 48149 M\"unster, Germany}
\affiliation{Center for Data Science and Complexity (CDSC),
University of M\"{u}nster, Corrensstrasse 2,  48149 M\"{u}nster, Germany}

\author{Uwe Thiele}
\email{\textcolor{blue}{u.thiele@uni-muenster.de}}
\homepage{http://www.uwethiele.de}
\thanks{ORCID ID: 0000-0001-7989-9271}
\affiliation{Institute of Theoretical Physics, University of M\"unster, Wilhelm-Klemm-Str.\ 9, 48149 M\"unster, Germany}
\affiliation{Center for Data Science and Complexity (CDSC),
University of M\"{u}nster, Corrensstrasse 2,  48149 M\"{u}nster, Germany}
\affiliation{Center for Multiscale Theory and Computation (CMTC), University of M\"unster, Corrensstr.\ 40, 48149 M\"unster, Germany}

\renewcommand{\abstractname}{}

\begin{abstract}
\vspace{1cm}
We derive and analyze a mean-field theory of the chiral Ising model recently introduced by Wang, Pietzonka, and Jülicher in \emph{Edge Currents Shape Condensates in Chiral Active Matter}, \href{https://arxiv.org/abs/2603.20064}{arXiv:2603.20064}. Starting from the master equation for clockwise and counterclockwise rotations of $2\times2$ spin blocks, we first obtain spatially discrete evolution equations for the spatially resolved average magnetization. On this discrete level, we show that a chiral bias strongly affects phase coarsening: domains coarsen anisotropically, develop nearly rectangular shapes, and eventually display chirality-induced arrested coarsening. Taking the continuum limit of these equations yields an active field theory that has the structure of a relaxational Model-B-type dynamics supplemented by a chiral current that permanently drives the system out of equilibrium. The coarse graining explicitly shows how microscopic rotational bias generates tangential currents localized at interfaces. Using this continuum theory, we perform a linear stability analysis of radially symmetric clusters and identify a 
chiral fingering instability in which angular perturbations of the interface are amplified and eventually lead to radially asymmetric rotating states or disordered states.
\end{abstract}

\maketitle
\newpage
\tableofcontents
\section{Introduction}
Chirality---the property of handedness or the lack of mirror symmetry---is a ubiquitous feature of nature that manifests itself across a vast range of scales. In organic chemistry and biology, chirality is fundamental: amino acids are almost exclusively left-handed \cite{Meierhenrich2008}, DNA winds in a right-handed double helix \cite{alberts2014molecular}, and many microorganisms swim along helical trajectories due to the chiral geometry of their flagella~\cite{Lauga_2009}. In physics, chirality plays a central role, from parity violation in the weak nuclear force~\cite{Griffiths:1987tj} to topological edge states in quantum Hall systems~\cite{RevModPhys.89.025005}. In soft and active matter, microscopic handedness can be amplified into macroscopic structure, transport, and collective motion \cite{Liebchen_2022,chiral_review}. Examples include rotating crystals and self-sustained chiral oscillations in rotating starfish embryos~\cite{Fakhri_starfish}, cholesteric order and blue phases in liquid crystals~\cite{degennes1993physics}, and emergent edge currents in active spinner materials~\cite{doi:10.1073/pnas.1609572113}. Across these soft- and active-matter examples, the common thread is that chirality introduces a preferred sense of rotation that is amplified from the microscopic scale to the collective level, making it a robust organizing principle for emergent nonequilibrium phenomena.

A particularly important class of active chiral systems is formed by active spinner or rotor models, in which particles are driven by a persistent torque rather than, or in addition to, self-propulsion. Continuum theories of such systems have shown that local torque generation can produce antisymmetric stresses, spin--flow coupling, odd viscosity, odd diffusion, and related transverse transport phenomena \cite{Furthauer2012Active,banerjee2017,Han2021Fluctuating,PhysRevFluids.7.043301,tkbx-75zt,marconi2026emergenthydrodynamicschiralactive,maity2025activechiralrotorshydrodynamics,doi:10.1073/pnas.2201279119}. These hydrodynamic descriptions are closely related to broader theories of odd transport and odd mechanical response in active materials \cite{annurev:/content/journals/10.1146/annurev-conmatphys-040821-125506,Scheibner2020,PhysRevLett.126.248001}.

At the particle level, simulations of torque-driven hard particles and self-spinning disks have shown that rotational activity alone can generate effective alignment, phase separation, rotating clusters, vortex-like collective motion, and rotating crystallites \cite{PhysRevLett.112.075701,dbgp-pqsh,PhysRevLett.125.218002,D4SM01426J,Caprini2024Self}. Closely related models of self-spinning dimers and active spinner materials exhibit active melting, spatiotemporal order, and robust edge currents whose direction is fixed by the microscopic handedness of the rotors \cite{doi:10.1073/pnas.1609572113,PhysRevE.101.022603,PhysRevE.108.014609}. Mixtures of oppositely rotating particles further enrich this phenomenology by producing spin-pairing, active polymerization, chiral flows, and flocking \cite{Gelvan2024Hydrodynamic,PhysRevResearch.4.033230,10.3389/fphy.2022.972051,10.1063/5.0135233,Kreienkamp_2022}. Collective chiral motion can also appear as an emergent phenomenon in microscopically achiral systems, for example as rotating crystallites \cite{m3dy-53yc}.

Experiments on granular spinners, magnetically or electrically driven colloidal rotors, Quincke rollers, acoustically powered microspinners, and active chiral fluids have confirmed many of these theoretical findings. Such systems display oscillatory collective motion, boundary-localized transport, cargo transport assisted by odd viscosity, hyperuniform chiral states, and reconfigurable chiral patterns \cite{PhysRevLett.94.214301,C8SM00402A,doi:10.1073/pnas.1922633117,PhysRevLett.126.198001,PhysRevLett.128.218002,Zhang2020Reconfigurable,Katuri2024Control,10.1063/5.0210859,McNeill2023Three,Ning2024Macroscopic}. Recent studies have also explored chiral active fluids in confinement, external potentials, complex geometries, and intermediate-Reynolds-number regimes, where boundaries and hydrodynamic interactions can strongly modify the resulting chiral flows \cite{PhysRevResearch.5.023196,D3SM00793F,Chen2024Self,9kfb-c51g,DELMOTTE2024113321,chatzittofi2026selfphoreticcolloidschiralactive}. 

Taken together, these works establish active spinners as a minimal and versatile setting in which microscopic chirality is converted into macroscopic chiral transport, interfacial currents, rotating phases, and odd mechanical response. Most of these examples, however, involve explicit particle motion or hydrodynamic flow fields. A complementary route is to ask how microscopic chirality is coarse-grained into particular features of a conserved scalar order parameter. For instance, a first-principles field theory has recently been derived for active chiral hard disks, providing a microscopic route from chiral particle dynamics to continuum equations such as active Model~B+ \cite{Kalz_2024}. Similarly, for lattice spin systems, we have recently shown that the mean-field approximation can systematically connect microscopic nonequilibrium spin dynamics to continuum active field theories, including the nonreciprocal Allen--Cahn and Cahn--Hilliard equations for achiral but nonreciprocally interacting spins \cite{10.21468/SciPostPhys.20.1.005}. Here, we pursue a similar program for the recently introduced chiral Ising model \cite{wang2026edgecurrentsshapecondensates}: starting from the microscopic $2\times 2$ block-rotation dynamics of Ising spins, we derive from the master equation a mean-field continuum description, which can be interpreted as a mean-field version of chiral active Model~B.

This paper proceeds in three steps: First, in Sec.~\ref{sec:energy} we formulate the microscopic chiral Ising model originally introduced in \cite{wang2026edgecurrentsshapecondensates} and derive in Sec.~\ref{sec:avg} an exact evolution equation for the spatially resolved average magnetization from the master equation. Applying the mean-field approximation in Sec.~\ref{sec:MF} yields closed discrete lattice equations, which show that chiral block rotations qualitatively modify coarsening dynamics by producing rectangular morphologies and ultimately arrested coarsening. We then take the continuum limit of these equations in Sec.~\ref{sec:cont} and obtain an active field theory for the conserved magnetization with a passive Cahn-Hilliard contribution and a chiral interfacial current generated by the microscopic rotational bias. Finally, we use the continuum theory to analyze the stability of radially symmetric steady states and show that sufficiently strong chiral driving induces a finite-wavelength angular instability of the interface. This chiral fingering instability provides a route from stable circular domains to either radially asymmetric rotating states or irregular, disordered morphologies.
\section{The Chiral Ising model}\label{sec:energy}
\subsection{Lattice setup and energetics}
\begin{figure}[t!]
    \centering\includegraphics[width=\linewidth]{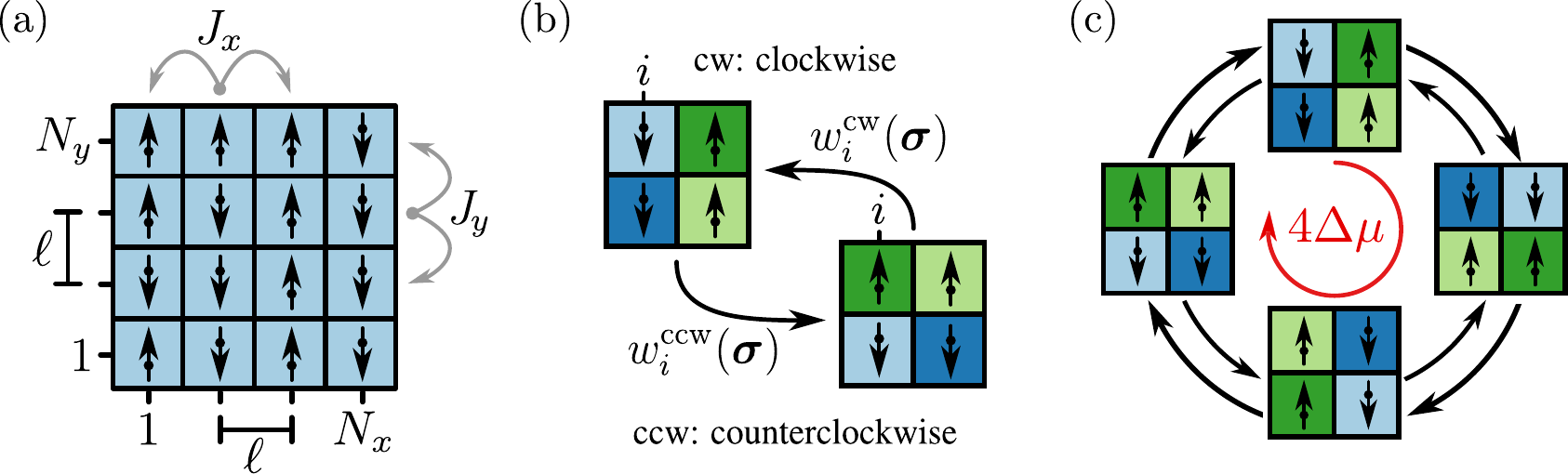}
    \caption{\textbf{2$\times$2 chiral block-rotation dynamics in the Ising model.}
(a) Schematic of the square lattice with lattice spacing $\ell$, lattice size $\{N_x,N_y\}$, and periodic boundary conditions; nearest-neighbor couplings are denoted by $J_x$ and $J_y$.
(b) Dynamics proceed via $2\times2$ block rotations in either the clockwise (cw) or counterclockwise (ccw) direction, with transition rates $w^{\rm cw}_{i}$ and $w^{\rm ccw}_{i}$, respectively, given by Eqs.~\eqref{Glauberratecw}. The different colors are used to visualize the block rotation.
(c) When the attempt probabilities $p$ and $1-p$ for cw and ccw rotations are unequal, the dynamics generate a nonzero steady-state cycle current proportional to the driving force $\Delta\mu=\ln{(p/[1-p])}$ [see Eq.~\eqref{f}].
}
    \label{Fig1}
\end{figure}

We consider a square lattice with periodic boundary conditions as shown in Fig.~\ref{Fig1}a. The lattice contains $N = N_{x}N_{y}$ spins with lattice spacing $\ell$, and each spin can attain values $\sigma_{i} = \pm1$ with $i \in \{1, \dots, N\}$ indexing its site. The local interaction energy of the spin located at lattice index $i$ is given by
\begin{equation}
    E_{i} = -\sigma_{i}h_{i}, 
    \label{Eh}
\end{equation}  
where the local field $h_i$ at lattice site $i$ is defined as
\begin{equation}
    h_i
    \equiv
    \sum_{\langle ij\rangle} J_{ij}\sigma_j,
    \label{h}
\end{equation}
with $\langle ij \rangle$ indicating a sum over the nearest-neighbor indices $j$ of lattice index $i$, and
\begin{equation}
    J_{ij} = J_{ji}
    =
    \begin{dcases}
        J_x, & j=i\pm\hat{x},\\
        J_y, & j=i\pm\hat{y},
    \end{dcases}
\end{equation}
is the interaction strength in the horizontal ($J_x$) and vertical ($J_y$) direction, respectively. The isotropic case is recovered by setting $J_x=J_y=J$. Throughout this work, energies are expressed in units of $k_{\rm B}T$, where $k_{\rm B}$ is the Boltzmann constant and $T$ denotes the temperature of the external heat bath (i.e., the thermostat) in which the system is immersed. Summing Eq.~\eqref{Eh} over all spins, we obtain the total energy for a specific configuration $\boldsymbol{\sigma}=\{\sigma_{1},\sigma_{2},\dots,\sigma_{N}\}$:
\begin{equation}
    \mathcal{H}(\boldsymbol{\sigma})\equiv\frac{1}{2}\sum_iE_i=-\frac{1}{2}\sum_i \sum_{\langle ij\rangle}J_{ij}\sigma_{i}\sigma_{j},
    \label{Hamiltonian}
\end{equation}
which is the canonical Ising model Hamiltonian \cite{ising1925beitrag}.
\subsection{Master equation for block rotations}\label{sec:model}
Our microscopic derivation of a continuum model starts from the master equation for $2\times2$ spin block rotation dynamics. To lighten the notation we use the superscripts $\rm cw$ and $\rm ccw$ to denote clockwise and counterclockwise rotations of a $2\times2$ block, respectively. Let $ P(\boldsymbol{\sigma};t)\in [0,1]$ denote the probability of finding the system in state  $\boldsymbol{\sigma}$ at time $t$. This probability evolves according to the master equation  
\begin{equation}
\frac{{\rm d}P(\boldsymbol{\sigma};t)}{{\rm d}t}{=}\sum_{i}\left[w^{\rm ccw}_i(\boldsymbol{\sigma}^{\rm cw}_{i})P(\boldsymbol{\sigma}^{\rm cw}_{i};t){+}w^{\rm cw}_i(\boldsymbol{\sigma}^{\rm ccw}_{i})P(\boldsymbol{\sigma}^{\rm ccw}_{i};t)
   {-} (w^{\rm cw}_i(\boldsymbol{\sigma}){+}w^{\rm ccw}_i(\boldsymbol{\sigma}))P(\boldsymbol{\sigma};t)\right],
   \label{masterequation}
\end{equation}
where  
$\boldsymbol{\sigma}^{\rm cw}_{i}$ and $\boldsymbol{\sigma}^{\rm ccw}_{i}$ are the configurations obtained from configuration $\boldsymbol{\sigma}$ by applying a clockwise (cw) or counterclockwise (ccw) rotation to the $2\times2$ block whose upper-left corner is lattice site $i$ (see Fig.~\ref{Fig1}b). Since a block rotation conserves the total magnetization, the configuration space $\Omega$ is of size
\begin{equation*}
|\Omega |=\binom{N}{N_{\uparrow}},    
\end{equation*}
where $N_{\uparrow}$ is the total number of up spins determined by the initial condition. For the considered $2\times2$ block-rotation dynamics, ergodicity within each fixed-magnetization sector has been shown in \cite{wang2026edgecurrentsshapecondensates}. Thus, for a given value of $N_{\uparrow}$, the dynamics can in principle visit all configurations in $\Omega$.
\subsection{Transition Rates}
The transition rates $w^{\rm cw}_i(\boldsymbol{\sigma})$ and $w^{\rm ccw}_i(\boldsymbol{\sigma})$ that enter Eq.~\eqref{masterequation} rotate a $2\times 2$ block with lattice index $i$ in the upper left corner (see Fig.~\ref{Fig1}b). We choose these rates to have a Glauber-type expression and read \cite{glauber_timedependent_1963}
\begin{align}
\begin{split}
    w^{\rm cw}_i(\boldsymbol{\sigma})&=\ \ \ \frac{p}{2\tau}  \  \left[1-\tanh{\left(\frac{\mathcal{H}(\boldsymbol{\sigma}^{\rm cw}_{i})-\mathcal{H}(\boldsymbol{\sigma})}{2}\right)}\right],\\
    w^{\rm ccw}_i(\boldsymbol{\sigma})&=\frac{1-p}{2\tau} \left[1-\tanh{\left(\frac{\mathcal{H}(\boldsymbol{\sigma}^{\rm ccw}_{i})-\mathcal{H}(\boldsymbol{\sigma})}{2}\right)}\right], 
\end{split}
\label{Glauberratecw}
\end{align}
where $p\in[0,1]$ determines the attempt probability for a clockwise rotation, and $\tau>0$ is an intrinsic timescale for a block rotation. For $p=1/2$ the attempt probabilities for a clockwise and counterclockwise rotation are equal, and for $p\neq 1/2$ there is a preferential direction of rotation. 
\subsection{Local Detailed Balance}\label{sec:DB}
The transition rates follow the thermodynamic constraint of \emph{local detailed balance} w.r.t.~the interaction energy given by Eq.~\eqref{Hamiltonian}. To see this, we use Eq.~\eqref{Glauberratecw} to compute the ratio of forward to backward transition rates from a starting configuration $\boldsymbol{\sigma}$:
\begin{align}
\begin{split}
    \frac{w^{\rm cw}_i(\boldsymbol{\sigma})}{w^{\rm ccw}_i(\boldsymbol{\sigma}^{\rm cw}_{i})}&=\frac{p}{1-p}\frac{1-\tanh{\left([\mathcal{H}(\boldsymbol{\sigma}^{\rm cw}_{i})-\mathcal{H}(\boldsymbol{\sigma})]/2\right)}}{1-\tanh{\left([\mathcal{H}(\boldsymbol{\sigma})-\mathcal{H}(\boldsymbol{\sigma}^{\rm cw}_{i}]/2\right)}}=\exp(-[\mathcal{H}(\boldsymbol{\sigma}^{\rm cw}_{i}){-}\mathcal{H}(\boldsymbol{\sigma})]+\Delta \mu), \\
    \frac{w^{\rm ccw}_i(\boldsymbol{\sigma})}{w^{\rm cw}_i(\boldsymbol{\sigma}^{\rm ccw}_{i})}&=\frac{1-p}{p}\frac{1-\tanh{\left([\mathcal{H}(\boldsymbol{\sigma}^{\rm ccw}_{i})-\mathcal{H}(\boldsymbol{\sigma})]/2\right)}}{1-\tanh{\left([\mathcal{H}(\boldsymbol{\sigma})-\mathcal{H}(\boldsymbol{\sigma}^{\rm ccw}_{i}]/2\right)}}=\exp(-[\mathcal{H}(\boldsymbol{\sigma}^{\rm ccw}_{i}){-}\mathcal{H}(\boldsymbol{\sigma})]-\Delta \mu),
\end{split}
\label{locdetbal}
\end{align}
where we have identified a nonequilibrium driving force given by 
\begin{equation}
    \Delta \mu\equiv \ln{\left(\frac{p}{1-p}\right)},
    \label{f}
\end{equation}
which is identical to the driving force in \cite{wang2026edgecurrentsshapecondensates}.
The sign of $\Delta\mu$ determines the preferred direction of rotation: $\Delta\mu>0$ favors clockwise rotations, while $\Delta\mu<0$ favors counterclockwise rotations. We can interpret $|\Delta\mu|$ as the thermodynamic cost required to maintain this rotational bias. In particular, $|\Delta\mu|=0$ for $p=1/2$, while $|\Delta\mu|\rightarrow\infty$ as $p\rightarrow0$ or $p\rightarrow1$, corresponding to unidirectional rotation. A fundamental cycle is obtained by applying four unidirectional rotations to the same block, which returns the system to its original spin configuration (see Fig.~\ref{Fig1}c). Along such a closed cycle the total change in interaction energy vanishes, while the nonequilibrium driving contributions add up. The corresponding cycle affinity is therefore $4\Delta\mu$ for four clockwise rotations and $-4\Delta\mu$ for four counterclockwise rotations.

For $p=1/2$, Eq.~\eqref{locdetbal} reduces to the detailed balance condition w.r.t.~the Hamiltonian $\mathcal H(\boldsymbol{\sigma})$. In this case the forward and backward transition rates satisfy
\begin{equation}
    w^{\rm cw}_i(\boldsymbol{\sigma})\pi(\boldsymbol{\sigma})
    =
    w^{\rm ccw}_i(\boldsymbol{\sigma}^{\rm cw}_{i})
    \pi(\boldsymbol{\sigma}^{\rm cw}_{i}),
    \qquad
    w^{\rm ccw}_i(\boldsymbol{\sigma})\pi(\boldsymbol{\sigma})
    =
    w^{\rm cw}_i(\boldsymbol{\sigma}^{\rm ccw}_{i})
    \pi(\boldsymbol{\sigma}^{\rm ccw}_{i}),
    \label{detbal}
\end{equation}
where $\pi(\boldsymbol{\sigma})$ is the Boltzmann-Gibbs distribution $
    \pi(\boldsymbol{\sigma})
    \propto\exp[-\mathcal H(\boldsymbol{\sigma})]$. Therefore, in the absence of nonequilibrium driving, the steady state of the master equation is the equilibrium Boltzmann-Gibbs distribution\footnote{To see this, we note that in the steady state the right-hand side of Eq.~\eqref{masterequation} must vanish. One possible way for the right-hand side to vanish is when the summand for fixed $i$ vanishes, i.e.,
\begin{equation*}
     w^{\rm ccw}_i(\boldsymbol{\sigma}^{\rm cw}_{i})P^{\rm eq}_{\rm s}(\boldsymbol{\sigma}^{\rm cw}_{i}){+}w^{\rm cw}_i(\boldsymbol{\sigma}^{\rm ccw}_{i})P^{\rm eq}_{\rm s}(\boldsymbol{\sigma}^{\rm ccw}_{i})
    {-} (w^{\rm cw}_i(\boldsymbol{\sigma}){+}w^{\rm ccw}_i(\boldsymbol{\sigma}))P^{\rm eq}_{\rm s}(\boldsymbol{\sigma})=0.
 \end{equation*}
 This condition is automatically fulfilled when we insert Eq.~\eqref{Peqs} for $P^{\rm eq}_{\rm s}$ and finally use Eq.~\eqref{detbal}.},
\begin{equation}
    \left.\lim_{t\to\infty}P(\boldsymbol{\sigma};t)\right|_{\Delta\mu=0}
    \equiv 
    P^{\rm eq}_{\rm s}(\boldsymbol{\sigma})
    =
    \pi(\boldsymbol{\sigma}).
    \label{Peqs}
\end{equation}
Note that the transition rates given by Eq.~\eqref{Glauberratecw} differ from the Metropolis rates employed in \cite{wang2026edgecurrentsshapecondensates}. In the unbiased case, $p=1/2$, both choices satisfy detailed balance w.r.t.~the same Ising Hamiltonian and therefore yield the same equilibrium steady state. However, for $p\neq 1/2$, the dynamics is permanently driven out of equilibrium, and the nonequilibrium steady state is no longer determined by the Hamiltonian alone. Instead, it may also depend on the kinetic rule, i.e.,~on the specific form of the transition rates.  
\section{Dynamics of the Average Magnetization}\label{sec:avg}
From the master equation given by Eq.~\eqref{masterequation} we can obtain an evolution equation for the spatially resolved average magnetization, defined as
\begin{equation}
    m_j(t) \equiv \sum_{\boldsymbol{\sigma}}\sigma_j(\boldsymbol{\sigma}) P(\boldsymbol{\sigma};t),
\end{equation}
where the notation $\sigma_j(\boldsymbol{\sigma})$ indicates that the value of $\sigma_j$ depends on the configuration $\boldsymbol{\sigma}$. Multiplying the left- and right-hand sides of Eq.~\eqref{masterequation} by $\sigma_j(\boldsymbol{\sigma})$ and summing over all states, we obtain:
\begin{align*}
\frac{{\rm d} m_j (t)}{{\rm d}t}
&{=}\sum_{\boldsymbol{\sigma}}\sigma_{j}(\boldsymbol{\sigma})\frac{{\rm d}P(\boldsymbol{\sigma};t)}{{\rm d}t}\\
&{\stackrel{\eqref{masterequation}}{=}}\sum_{\boldsymbol{\sigma}}\sigma_{j}(\boldsymbol{\sigma})\!\sum_{i}\!\left[
w^{\rm ccw}_i(\boldsymbol{\sigma}^{\rm cw}_{i})P(\boldsymbol{\sigma}^{\rm cw}_{i};t)
{+}w^{\rm cw}_i(\boldsymbol{\sigma}^{\rm ccw}_{i})P(\boldsymbol{\sigma}^{\rm ccw}_{i};t)
{-}(w^{\rm cw}_i(\boldsymbol{\sigma}){+}w^{\rm ccw}_i(\boldsymbol{\sigma}))P(\boldsymbol{\sigma};t)
\right]\\
&{=}
\sum_{i}\!\sum_{\boldsymbol{\sigma}}\!\sigma_{j}(\boldsymbol{\sigma})[
\underbrace{
  w^{\rm ccw}_i(\boldsymbol{\sigma}^{\rm cw}_{i})P(\boldsymbol{\sigma}^{\rm cw}_{i};t)
  {+} 
  w^{\rm cw}_i(\boldsymbol{\sigma}^{\rm ccw}_{i})P(\boldsymbol{\sigma}^{\rm ccw}_{i};t)
}_{\text{gain}}
\!{-}\!
\underbrace{
  (w^{\rm cw}_i(\boldsymbol{\sigma}){+}w^{\rm ccw}_i(\boldsymbol{\sigma}))
  P(\boldsymbol{\sigma};t)
}_{\text{loss}}
],
\end{align*}
where in the last step we have interchanged the sums over the lattice index $i$ and the configurations $\boldsymbol{\sigma}$. 
The loss term on the right-hand side may be written directly as
\begin{equation*}
\sum_{\boldsymbol{\sigma}}\sigma_{j}(\boldsymbol{\sigma})(w^{\rm cw}_i(\boldsymbol{\sigma})+w^{\rm ccw}_i(\boldsymbol{\sigma}))P(\boldsymbol{\sigma};t)
\equiv \langle \sigma_{j}(w^{\rm cw}_i+w^{\rm ccw}_i)\rangle.
\end{equation*}
To handle the gain terms, it is convenient to introduce the rotation operators
\begin{equation*}
R_i^{\rm cw}\boldsymbol{\sigma}\equiv \boldsymbol{\sigma}^{\rm cw}_i,
\qquad
R_i^{\rm ccw}\boldsymbol{\sigma}\equiv \boldsymbol{\sigma}^{\rm ccw}_i ,
\end{equation*}
i.e.\ $R_i^{\rm cw}$ rotates the $2\times2$ block with upper-left index $i$ clockwise, and $R_i^{\rm ccw}$ rotates it counterclockwise.  
The two maps are bijections that are inverse to each other:
\begin{equation*}
\bigl(R_i^{\rm cw}\bigr)^{-1}=R_i^{\rm ccw},\qquad
\bigl(R_i^{\rm ccw}\bigr)^{-1}=R_i^{\rm cw}.
\end{equation*}
Consider for example the gain term
\begin{equation*}
\sum_{\boldsymbol{\sigma}}\sigma_{j}(\boldsymbol{\sigma})w^{\rm ccw}_i(\boldsymbol{\sigma}^{\rm cw}_{i})P(\boldsymbol{\sigma}^{\rm cw}_{i};t).
\end{equation*}
We now make the change of variables
\begin{equation*}
\boldsymbol{\sigma}' \equiv R_i^{\rm cw}\boldsymbol{\sigma} = \boldsymbol{\sigma}^{\rm cw}_i \qquad \Longleftrightarrow \qquad \boldsymbol{\sigma} = R_i^{\rm ccw}\boldsymbol{\sigma}'.
\end{equation*}
Since $R_i^{\rm cw}$ is a bijection, summing over $\boldsymbol{\sigma}$ equals summing over $\boldsymbol{\sigma}'$:
\begin{align*}
\sum_{\boldsymbol{\sigma}}\sigma_{j}(\boldsymbol{\sigma})w^{\rm ccw}_i(\boldsymbol{\sigma}^{\rm cw}_{i})P(\boldsymbol{\sigma}^{\rm cw}_{i};t)
&=\sum_{\boldsymbol{\sigma}'}\sigma_{j}\left(R_i^{\rm ccw}\boldsymbol{\sigma}'\right)w^{\rm ccw}_i(\boldsymbol{\sigma}')P(\boldsymbol{\sigma}';t).
\end{align*}
Now, the only difference between $\boldsymbol{\sigma}'$ and $R_i^{\rm ccw}\boldsymbol{\sigma}'$ is in the rotated $2\times2$ block with index $i$ in its upper left corner.  
Thus:
\begin{equation*}
\sum_{\boldsymbol{\sigma}}\sigma_{j}(\boldsymbol{\sigma})w^{\rm ccw}_i(\boldsymbol{\sigma}^{\rm cw}_{i})P(\boldsymbol{\sigma}^{\rm cw}_{i};t)=
\begin{dcases}
\langle \sigma_{j}w^{\rm ccw}_i \rangle, & i\notin\mathcal{A}_j,\\[4pt]
\langle \sigma_{\pi_i^{\rm ccw}(j)}w^{\rm ccw}_i \rangle, & i\in\mathcal{A}_j,
\end{dcases}
\end{equation*}
where $\mathcal{A}_j$ is the influence set of lattice indices whose block rotation affects the spin value at index $j$ (see Fig.~\ref{Fig2}a for an example), and
$\pi_i^{\rm ccw}(j)$ denotes the index inside the same block whose spin moves to lattice index $j$ under a counterclockwise rotation. Analogously,
\begin{equation*}
\sum_{\boldsymbol{\sigma}}\sigma_{j}(\boldsymbol{\sigma})w^{\rm cw}_i(\boldsymbol{\sigma}^{\rm ccw}_{i})P(\boldsymbol{\sigma}^{\rm ccw}_{i};t)=
\begin{dcases}
\langle \sigma_{j}w^{\rm cw}_i \rangle, & i\notin\mathcal{A}_j,\\[4pt]
\langle \sigma_{\pi_i^{\rm cw}(j)}w^{\rm cw}_i \rangle, & i\in\mathcal{A}_j.
\end{dcases}
\end{equation*}
Putting all terms back into the expression for ${\rm d} m_j/{\rm d}t$, one finds that all contributions with $i\notin\mathcal{A}_j$ cancel (gain $=$ loss). Hence,
\begin{equation}
\frac{{\rm d} m_j (t)}{{\rm d}t}
=
\sum_{i\in\mathcal{A}_j}
\langle
(\sigma_{\pi_i^{\rm cw}(j)}-\sigma_j)w^{\rm cw}_i
+
(\sigma_{\pi_i^{\rm ccw}(j)}-\sigma_j)w^{\rm ccw}_i
\rangle.
\label{finalresult}
\end{equation}
So, the time-evolution of the magnetization at site $j$ is given by the expected change of $\sigma_j$ induced by all $2\times2$ rotations that involve site $j$. A specific example of this will be given in the next section by Eq.~\eqref{Eqm0MF}.  Note that Eq.~\eqref{finalresult} is not yet in closed form, as the right-hand side takes the form of an average over the nonlinear transition rates $w^{\rm cw}_i$ and $w^{\rm ccw}_i$. To get a closed-form expression, we will apply the MF approximation in the next section. 
\section{Mean-field approximation}\label{sec:MF}
\subsection{Introducing the mean-field approximation}\label{sec:MFnote}
The MF approximation consists of two steps. First, when evaluating the expectation value in Eq.~\eqref{finalresult}, we approximate the full probability distribution $P(\boldsymbol{\sigma};t)$ by a product measure of independent one-site Bernoulli probabilities \cite{Ducastelle1996}. Thus, correlations between different lattice sites are neglected, while the local mean spin values $m_i=\langle\sigma_i\rangle$ are retained. More explicitly, for any nonlinear function $f=f(\sigma_{i_1},\dots,\sigma_{i_n})$ depending on the spins $\sigma_{i_1},\dots,\sigma_{i_n}$, we use
\begin{align}
\langle f(\sigma_{i_1},\dots,\sigma_{i_n})\rangle
\stackrel{\rm MF}{=}
\sum_{\sigma_{i_1}=\pm1}\cdots\sum_{\sigma_{i_n}=\pm1}
\left[
\prod_{k=1}^n p_{\rm MF}(\sigma_{i_k}|m_{i_k})
\right]
f(\sigma_{i_1},\dots,\sigma_{i_n}),
\label{MFsumgeneral}
\end{align}
where
\begin{equation}
p_{\rm MF}(\sigma_i|m_i)\equiv
\frac{1+\sigma_i m_i}{2},
\qquad
\sigma_i=\pm1,
\label{BernoulliMFgeneral}
\end{equation}
is the one-site Bernoulli probability associated with the local average spin $m_i$. This form for the probability is fixed by the requirement that it is normalized and has first moment $m_i$, i.e.,
\begin{align*}
\sum_{\sigma_i=\pm1}p_{\rm MF}(\sigma_i|m_i)=1,
\qquad
\sum_{\sigma_i=\pm1}\sigma_i p_{\rm MF}(\sigma_i|m_i)=m_i .
\end{align*}
In this way, the MF approximation replaces all higher-order spin correlations by products of local one-site averages, while preserving the prescribed local magnetization field. 

Second, whenever the local field $h_i$ appears, we replace the fluctuating field by its average value\footnote{A possible refinement would be to keep the local fields $h_i$ as fluctuating quantities and to include the neighboring spins entering $h_i$ in the Bernoulli product average. This would still neglect spatial correlations, but it would retain local-field fluctuations instead of replacing them by their mean value. Such a closure goes beyond the MF approximation used here and is closely related to effective-field approximations of Ising-type models \cite{PhysRevE.111.024207}.},
\begin{equation}
    h_i \stackrel{\rm MF}{=} \langle h_i \rangle 
    =  \sum_{\langle ij \rangle} J_{ij}m_j.
    \label{mf1}
\end{equation}
These two ingredients define the closure used in the following derivation.
\begin{figure}
    \centering
    \includegraphics[width=\textwidth]{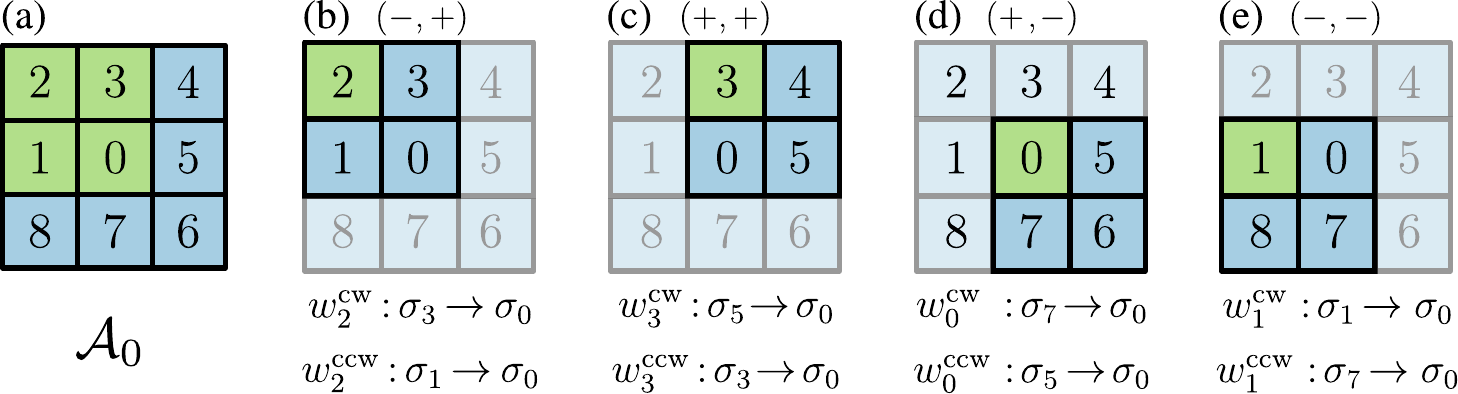}
    \caption{(a) The indices highlighted in green form the influence set $\mathcal{A}_0 = \{0,1,2,3\}$ of the spin value at site $0$. These are precisely the indices marking the top left site of $2\times 2$ blocks whose rotation can alter the spin at site $0$. Panels (b)–(e) illustrate how rotating each of the four indices in the influence set $\mathcal{A}_0=\{0,1,2,3\}$ affects the spin at site $0$. For each panel, a clockwise (cw) or counterclockwise (ccw) $2\times2$ block rotation is performed with the indicated  spin in the upper–left corner of the block. The resulting value of $\sigma_0$ after  the rotation is shown: for example, in panel (b), a cw rotation maps  $\sigma_3 \rightarrow \sigma_0$, whereas a ccw rotation maps  $\sigma_1 \rightarrow \sigma_0$. The notation $(-,+)$, $(+,+)$ etc.~above panel (b)-(e) indicates the center position of the block relative to lattice site 0.}
    \label{Fig2}
\end{figure}
\subsection{Applying the mean-field approximation}
 We now illustrate how to evaluate Eq.~\eqref{finalresult} using the MF approximation for the setup shown in Fig.~\ref{Fig2}. For concreteness, we derive the evolution equation for the average spin at lattice index $0$. Using Eq.~\eqref{finalresult}, we obtain
\begin{align}
\frac{{\rm d} m_0 (t)}{{\rm d}t}
&=
\langle
(\sigma_{1}-\sigma_0)(w^{\rm cw}_1+w^{\rm ccw}_2)
\rangle
+\langle
(\sigma_{3}-\sigma_0)(w^{\rm cw}_2+w^{\rm ccw}_3)
\rangle
\nonumber\\
&
+\langle
(\sigma_{5}-\sigma_0)(w^{\rm cw}_3+w^{\rm ccw}_0)
\rangle
+\langle
(\sigma_{7}-\sigma_0)(w^{\rm cw}_0+w^{\rm ccw}_1)
\rangle.
\label{Eqm0MF}
\end{align}
To better understand the structure of Eq.~\eqref{Eqm0MF}, let us consider the first term on the right-hand side. This term collects precisely the two block rotations that replace the spin located at index $0$ by the spin located at index $1$. These are (i) a clockwise rotation of the block whose upper-left corner is at lattice index $1$ (see Fig.~\ref{Fig2}e), and (ii) a counterclockwise rotation of the neighboring block whose upper-left corner is at lattice index $2$ (see Fig.~\ref{Fig2}b). Hence, both processes contribute to the change of $m_0$ through the factor $(\sigma_1-\sigma_0)$, which is simply the change of the spin value at site $0$ induced by these two allowed block rotations. For the non-interacting case $J{=}0$, the transition rates simplify considerably, and Eq.~\eqref{Eqm0MF} becomes
\begin{align*}
\left.\tau\frac{{\rm d} m_{0} (t)}{{\rm d}t}\right|_{J=0}
=\frac{1}{2}\langle\sigma_{1}+\sigma_{3}+\sigma_{5}+\sigma_{7}-4\sigma_{0}\rangle
=\frac{1}{2}\left(m_1+m_3+m_5+m_7-4m_0\right),
\end{align*}
which is the discrete Laplacian acting on $m_0$. Thus, random uncorrelated block rotations reduce to ordinary diffusion, independently of whether the rotations are biased clockwise or counterclockwise. For $J\neq0$, the structure of the right-hand side of Eq.~\eqref{Eqm0MF} becomes more involved. To make this explicit, we focus on the two rates $w^{\rm cw}_1$ and $w^{\rm ccw}_1$, which read
\begin{align}
    \begin{split}
    w^{\rm cw}_1
    &=
    p
    \left[
    1-\tanh\left(
    \Delta \mathcal H^{\rm cw}_1/2
    \right)
    \right]/2\tau,
    \\
    w^{\rm ccw}_1
    &=
    (1-p)
    \left[
    1-\tanh\left(
    \Delta \mathcal H^{\rm ccw}_1/2
    \right)
    \right]/2\tau,
    \end{split}
    \label{weff}
\end{align}
with
\begin{align}
    \Delta \mathcal H^{\rm cw}_1
    &\stackrel{\rm MF}{=}
    -
    \left[
    (\sigma_1-\sigma_0)\langle h_0 \rangle 
    +
    (\sigma_8-\sigma_1)\langle h_1 \rangle
    +
    (\sigma_7-\sigma_8) \langle h_8 
    \rangle
    +
    (\sigma_0-\sigma_7) \langle h_7 \rangle
    \right],
    \label{DeltaHcwLocal}
    \\
    \Delta \mathcal H^{\rm ccw}_1
    &\stackrel{\rm MF}{=}
    -
    \left[
    (\sigma_7-\sigma_0)\langle h_0 \rangle
    +
    (\sigma_0-\sigma_1)\langle h_1 \rangle
    +
    (\sigma_1-\sigma_8)\langle h_8 
    \rangle
    +
    (\sigma_8-\sigma_7)\langle h_7 \rangle
    \right].
    \label{DeltaHccwLocal}
\end{align}
Here the local field $h_i$ is defined in Eq.~\eqref{h}, and in Eqs.~\eqref{DeltaHcwLocal}--\eqref{DeltaHccwLocal} it has been replaced by its mean value $\langle h_i\rangle$ according to Eq.~\eqref{mf1}. Since the local fields also contain interactions from within the $2\times2$ block,  the energy differences entering the rates are \emph{not} the exact energy changes of a simultaneous $2\times2$ block rotation. Instead, they are evaluated using a local spin-replacement approximation. In this approximation, each site in the rotating block is treated separately: the spin arriving at site $i$ after the block rotation replaces the spin that was previously located at site $i$, and the corresponding local energy change is computed in the mean field $\langle h_i\rangle$. For example, in the clockwise rotation associated with $w^{\rm cw}_1$, the spin at site $0$ is replaced by $\sigma_1$, the spin at site $1$ by $\sigma_8$, the spin at site $8$ by $\sigma_7$, and the spin at site $7$ by $\sigma_0$. The four terms in Eq.~\eqref{DeltaHcwLocal} are precisely the corresponding local replacement contributions. The counterclockwise expression in Eq.~\eqref{DeltaHccwLocal} is obtained analogously. These expressions differ from the exact block-rotation energy, where energy changes coming from interactions inside the rotating $2\times2$ block are absent. In the local spin-replacement approximation used here, the mean fields $\langle h_i\rangle$ include all nearest neighbors of site $i$, including neighbors inside the same $2\times2$ block. Consequently, intrablock interactions contribute to the transition rates. A similar local-field approximation is commonly used when deriving model B-type continuum dynamics from spin-exchange dynamics \cite{Penrose1991,10.21468/SciPostPhys.20.1.005}. We now evaluate the corresponding expectation values using the MF approximation introduced in Sec.~\ref{sec:MFnote}. In particular, we use the product of one-site Bernoulli probabilities \eqref{BernoulliMFgeneral} for the four spins forming the rotating block. Thus,
\begin{align}
\begin{split}
\langle (\sigma_1-\sigma_0)w^{\rm cw}_1\rangle
&\stackrel{\rm MF}{=}
\sum_{\sigma_1=\pm1}\sum_{\sigma_0=\pm1}
\sum_{\sigma_7=\pm1}\sum_{\sigma_8=\pm1}
\left[
\prod_{i\in\{1,0,7,8\}} p_{\rm MF}(\sigma_i| m_i)
\right]
(\sigma_1-\sigma_0)w^{\rm cw}_1, \\
\langle (\sigma_7-\sigma_0)w^{\rm ccw}_1\rangle
&\stackrel{\rm MF}{=}
\sum_{\sigma_1=\pm1}\sum_{\sigma_0=\pm1}\sum_{\sigma_7=\pm1}\sum_{\sigma_8=\pm1}\left[
\prod_{i\in\{1,0,7,8\}} p_{\rm MF}(\sigma_i| m_i)
\right]
(\sigma_7-\sigma_0)w^{\rm ccw}_1.
\label{MFsum2}
\end{split}
\end{align}
The sums in Eqs.~\eqref{MFsum2} run over only $2^4=16$ spin configurations and can therefore be evaluated exactly. To separate the equilibrium contribution from the driven part, it is convenient to parametrize the rotation probability as
\begin{equation}
p\equiv \frac{1+\delta}{2},
\qquad \delta\in[-1,1],
\label{defdelta}
\end{equation}
where $\delta$ controls the strength and sign of the nonequilibrium driving. 
After carrying out the sums explicitly, the result naturally splits into a $\delta$-independent equilibrium part (denoted with subscript $\rm eq$) and a nonequilibrium contribution proportional to $\delta$ (denoted with subscript $\rm neq$):
\begin{align}
    &\left.8\tau\langle (\sigma_{1}-\sigma_0)w^{\rm cw}_1+(\sigma_{7}-\sigma_0)w^{\rm ccw}_1\rangle\right|_{{\rm  eq}}\stackrel{\rm MF}{=}2(m_1+m_7)-4m_0+\nonumber\\
    &(1{-}m_0m_1{-}m_0m_7{+}m_0m_8{+}m_1m_7{-}m_1m_8{-}m_7m_8{+}m_0m_1m_7m_8)\tanh{(\langle h_0\rangle{+}\langle h_8\rangle{-}\langle h_1\rangle{-}\langle h_7\rangle)}+\nonumber\\
    &(1-m_0m_8)(1-m_1m_7)\tanh{(\langle h_0\rangle-\langle h_8\rangle)}+\nonumber\\
    &(1-m_0m_7)(1+m_1m_8)\tanh{(\langle h_0\rangle-\langle h_7\rangle)}+\nonumber\\
    &(1-m_0m_1)(1+m_7m_8)\tanh{(\langle h_0\rangle -\langle h_1 \rangle )},
    \label{eqpart}
\end{align}
and
\begin{align}
    &\!\!\left.8\tau\langle (\sigma_{1}{-}\sigma_0)w^{\rm cw}_1{+}(\sigma_{7}{-}\sigma_0)w^{\rm ccw}_1\rangle\right|_{{\rm  neq}}{\stackrel{\rm MF}{=}}2\delta (m_1{-}m_7){-}
    \delta(m_0{-}m_8)(m_1{-}m_7)\tanh{(\langle h_0\rangle{-}\langle h_8\rangle)}{-}\nonumber\\ 
    &\delta(m_0-m_7)(m_1+m_8)\tanh{(\langle h_0\rangle-\langle h_7\rangle)}+\delta(m_0-m_1)(m_7+m_8)\tanh{(\langle h_0 \rangle-\langle h_1\rangle)}.
    \label{neqpart}
\end{align}
The corresponding contributions from the remaining three blocks entering Eq.~\eqref{Eqm0MF} are obtained analogously by lattice symmetry. Note that the first term in Eq.~\eqref{neqpart}, $2\delta (m_1-m_7)$, will eventually cancel with the terms coming from the remaining three blocks, and therefore only the terms containing $\tanh{}$ remain. Since Eq.~\eqref{eqpart} comes from the equilibrium ($\delta$-independent) part, we expect that one can bring it into a gradient dynamics form. This we do in the next section.  
\subsubsection{Equilibrium part and gradient dynamics}
 Here, we show how Eq.~\eqref{eqpart} can be cast into a gradient-dynamics form. The key step is to rewrite the first term on the right-hand side of Eq.~\eqref{eqpart} using the following exact identity:
\begin{align}
    2(m_1+m_7)-4m_0&=m_1+m_7-m_0-m_8+m_0m_1m_8+m_0m_7m_8-m_0m_1m_7-m_1m_7m_8\nonumber \\
    &+(m_8-m_0)(1-m_1m_7)\nonumber\\
    &+(m_7-m_0)(1+m_1m_8)\nonumber\\
    &+(m_1-m_0)(1+m_7m_8).
    \label{rewrite}
\end{align}
We then combine terms pairwise: the first term on the right-hand side of Eq.~\eqref{rewrite} is added to the contribution on the second line of Eq.~\eqref{eqpart}, the second term is added to the contribution on the third line of Eq.~\eqref{eqpart}, and so forth for the remaining terms. To rewrite each resulting combination in gradient form, we repeatedly employ the identity
\begin{equation}
    y+c\tanh{(x)}=(c+y\tanh{(x)})\tanh{(\arctanh{(y/c)}+x)}.
    \label{trick}
\end{equation}
This allows us to express each contribution in Eq.~\eqref{eqpart} as a nonnegative mobility multiplied by a hyperbolic tangent of a free-energy derivative. Omitting intermediate algebraic steps for brevity, the resulting expressions read
\begin{align}
    &(1{-}m_0m_1{-}m_0m_7{+}m_0m_8{+}m_1m_7{-}m_1m_8{-}m_7m_8{+}m_0m_1m_7m_8)\tanh{(\langle h_0\rangle{+}\langle h_8 \rangle{-}\langle h_1 \rangle {-}\langle h_7 \rangle)} \nonumber \\
    &{+}m_1{+}m_7{-}m_0{-}m_8{+}m_0[m_1m_8{+}m_7m_8{-}m_1m_7]{-}m_1m_7m_8=
    \mathcal{M}^{-,-}_{0,{\rm a}}\tanh{\left(\mu_1{+}\mu_7{-}\mu_0{-}\mu_8\right)}, \nonumber \\
   &(m_8{-}m_0)(1{-}m_1m_7){+}
   (1{-}m_0m_8)(1{-}m_1m_7)\tanh{(\langle h_0 \rangle{-}\langle h_8\rangle )}=
   \mathcal{M}^{-,-}_{0,{\rm d}}\tanh{\left(\mu_8-\mu_0\right)}, \label{final1}\\
    &(m_7{-}m_0)(1{+}m_1m_8){+}
    (1{-}m_0m_7)(1{+}m_1m_8)\tanh{(\langle h_0 \rangle {-}\langle h_7\rangle)}=\mathcal{M}^{-,-}_{0,y}\tanh{\left(\mu_7-\mu_0\right)}, \nonumber \\
    &(m_1{-}m_0)(1{+}m_7m_8){+}
    (1{-}m_0m_1)(1{+}m_7m_8)\tanh{(\langle h_0 \rangle{-}\langle h_1 \rangle)}=  \mathcal{M}^{-,-}_{0,x}\tanh{\left(\mu_1-\mu_0\right)}, \nonumber
\end{align}
where $\mu_i$ is the MF chemical potential at site $i$
\begin{equation}
    \mu_i\equiv \frac{\partial \mathcal{F}_{\rm MF}}{\partial m_i},
    \label{mu}
\end{equation}
with 
$\mathcal{F}_{\rm MF}=\mathcal{F}_{\rm MF}(\mathbf{m})$, $\mathbf{m}=\{m_{0},m_{1},\ldots\}$ the canonical MF free energy
\begin{equation}
    \mathcal{F}_{\rm MF}(\mathbf{m})\equiv\frac{1}{2}\sum_{i=1}^{N}{\Large[}\Phi(m_i)-m^{}_{i}\sum_{\langle ij \rangle}J_{ij}m_{j}{\Large]},
    \label{FMF}
\end{equation}
and the MF entropy function\footnote{Note that $$\frac{1}{2}\frac{{\rm d}\Phi(m_i)}{{\rm d}m_i}=\arctanh{(m_i)}, $$ which relates to the $\arctanh$ term in Eq.~\eqref{trick}.}
\begin{equation}
     \Phi(m_i)\equiv (1+m_i)\ln{(1+m_i)}+(1-m_i)\ln{(1-m_i)}.
     \label{MFentropy}
\end{equation}
The mobilities $\mathcal{M}$ appearing in Eqs.~\eqref{final1} carry the superscript $(-,-)$ to indicate that they correspond to the spin block located to the southwest of site~$0$ (see Fig.~\ref{Fig2}). They are given explicitly by
\begin{align}
    \begin{split}
    \mathcal{M}^{-,-}_{0,{\rm a}}&=(1-m_7m_8)\left[1-m_0m_1+(m_1-m_0)\tanh{(\langle h_0 \rangle +\langle h_8\rangle -\langle h_1\rangle -\langle h_7\rangle)}\right] \\
    &+(1-m_1m_8)\left[1-m_0m_7+(m_7-m_0)\tanh{(\langle h_0 \rangle +\langle h_8\rangle -\langle h_1\rangle -\langle h_7\rangle)}\right] \\
    &-(1-m_1m_7)\left[1-m_0m_8+(m_8-m_0)\tanh{(\langle h_0 \rangle +\langle h_8\rangle -\langle h_1\rangle -\langle h_7\rangle)}\right],\\
    \mathcal{M}^{-,-}_{0,{\rm d}}&=(1-m_1m_7)\left[1-m_0m_8+(m_8-m_0)\tanh{(\langle h_0\rangle-\langle h_8 \rangle )}\right], \\
    \mathcal{M}^{-,-}_{0,y}&=(1+m_1m_8)\left[1-m_0m_7+(m_7-m_0)\tanh{(\langle h_0\rangle-\langle h_7 \rangle )}\right], \\
    \mathcal{M}^{-,-}_{0,x}&=(1+m_7m_8)\left[1-m_0m_1+(m_1-m_0)\tanh{(\langle h_0\rangle-\langle h_1 \rangle )}\right].
    \end{split}
    \label{mobilities}
\end{align}
The subscripts $\{{\rm a},{\rm d},x,y\}$ distinguish the mobilities associated with free-energy differences involving, respectively, all spins in the block, the diagonal pair, and the horizontal and vertical directions. In Appendix~\ref{AppendixMobilities} we show that all mobilities in Eq.~\eqref{mobilities} are nonnegative.
\subsubsection{Combining all results}
 Upon including the contributions from the other transition rates entering Eq.~\eqref{Eqm0MF}, the final result for the evolution equation of $m_0$ reads (using the index labelling shown in Fig.~\ref{Fig2})
\begin{align}
    8\tau \frac{{\rm d}m_{0}}{{\rm d}t}&=\mathcal{M}^{-,+}_{0,{\rm a}}\tanh{\left(\mu_1+\mu_3-\mu_0-\mu_2\right)}
    +
    \mathcal{M}^{+,+}_{0,{\rm a}}\tanh{\left(\mu_3+\mu_5-\mu_0-\mu_4\right)}\nonumber \\
    &+
    \mathcal{M}^{+,-}_{0,{\rm a}}\tanh{\left(\mu_5+\mu_7-\mu_0-\mu_6\right)} 
    +\mathcal{M}^{-,-}_{0,{\rm a}}\tanh{\left(\mu_1+\mu_7-\mu_0-\mu_8\right)}
    \nonumber \\
    &+
    (\mathcal{M}^{-,-}_{0,x}+\mathcal{M}^{-,+}_{0,x})\tanh{\left(\mu_1-\mu_0\right)}
    +
    (\mathcal{M}^{+,+}_{0,x}+\mathcal{M}^{+,-}_{0,x})\tanh{\left(\mu_5-\mu_0\right)} \nonumber \\
    &+
    (\mathcal{M}^{-,-}_{0,y}+\mathcal{M}^{+,-}_{0,y})\tanh{\left(\mu_7-\mu_0\right)}
    +
    (\mathcal{M}^{-,+}_{0,y}+\mathcal{M}^{+,+}_{0,y})\tanh{\left(\mu_3-\mu_0\right)} \nonumber \\
    &+
    \mathcal{M}^{-,-}_{0,{\rm d}}\tanh{\left(\mu_8-\mu_0\right)}+
    \mathcal{M}^{-,+}_{0,{\rm d}}\tanh{\left(\mu_2-\mu_0\right)} \nonumber \\
    &+
    \mathcal{M}^{+,+}_{0,{\rm d}}\tanh{\left(\mu_4-\mu_0\right)} 
    +
    \mathcal{M}^{+,-}_{0,{\rm d}}\tanh{\left(\mu_6-\mu_0\right)} \nonumber \\
     &+\delta(m_0-m_8)(m_7-m_1)\tanh{(\langle h_0\rangle - \langle h_8 \rangle)}
    \nonumber \\
    &+\delta(m_0-m_2)(m_1-m_3)\tanh{(\langle h_0\rangle - \langle h_2 \rangle)} 
    \nonumber \\
    &+\delta(m_0-m_4)(m_3-m_5)\tanh{(\langle h_0\rangle - \langle h_4 \rangle)} 
    \nonumber \\
    &+\delta(m_0-m_6)(m_5-m_7)\tanh{(\langle h_0\rangle - \langle h_6 \rangle)} \nonumber \\
    &+\delta(m_0-m_7)(m_5+m_6-m_1-m_8)\tanh{(\langle h_0\rangle - \langle h_7 \rangle)} 
    \nonumber \\
    &+\delta(m_0-m_1)(m_7+m_8-m_2-m_3)\tanh{(\langle h_0\rangle - \langle h_1 \rangle)} 
    \nonumber \\
    &+\delta(m_0-m_3)(m_1+m_2-m_4-m_5)\tanh{(\langle h_0\rangle - \langle h_3 \rangle)} 
    \nonumber \\
    &+\delta(m_0-m_5)(m_3+m_4-m_6-m_7)\tanh{(\langle h_0\rangle - \langle h_5 \rangle)}.
    \label{dm0final}
\end{align}
Because the lattice dynamics is translationally invariant, Eq.~\eqref{dm0final} immediately generalizes to any lattice site $i$: one simply replaces the reference index $0$ by $i$ and relabels all surrounding indices according to the same local geometry. This results in a system of $N$ coupled ordinary differential equations that can be numerically solved, e.g.~by numerical time integration. 
\begin{figure}[t!]
    \centering
    \includegraphics[width=\linewidth]{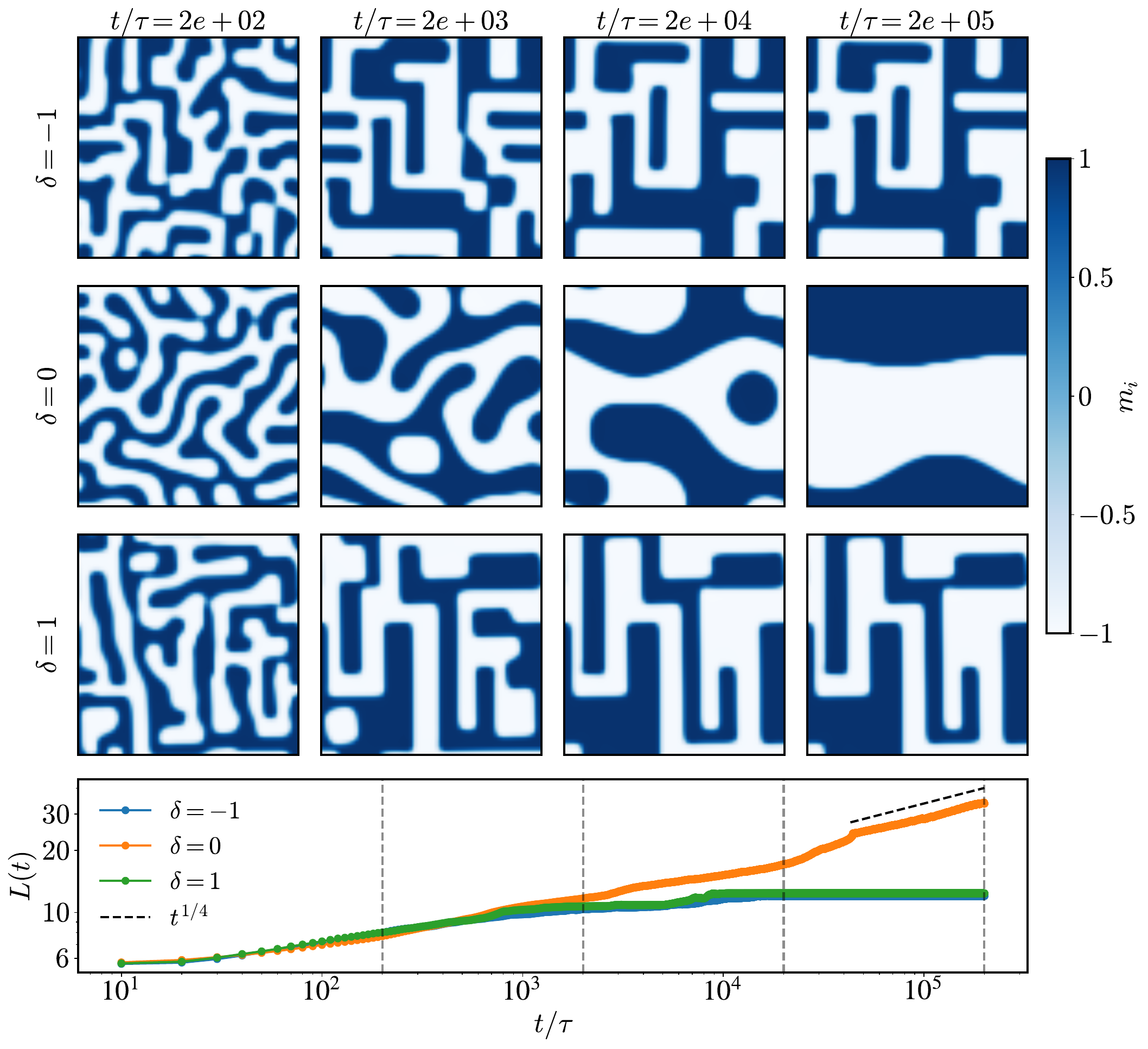}
    \caption{\textbf{Mean-field coarsening dynamics for different rotational biases.}
Time evolution of the MF lattice equations Eq.~\eqref{dm0final} on a $60\times60$ square lattice with periodic boundary conditions and isotropic coupling strength $J_x=J_y=0.6$. The upper three rows show snapshots of the magnetization field for different values of the chirality parameter: $\delta=-1$ in the upper row, $\delta=0$ in the middle row, and $\delta=1$ in the lower row. From left to right, the snapshots are taken at $t/\tau=2\times\{10^2,10^3,10^4,10^5\}$, as indicated by the vertical dashed lines in the lower panel. The lower panel shows the coarsening of the corresponding characteristic length $L(t)$ (see Appendix~\ref{app:lengthscale}). For $\delta=0$, the system undergoes continued phase separation, with a late-time growth law approximately consistent with $L(t)\sim t^{1/4}$. By contrast, for fully biased dynamics, $\delta=\pm1$, the domains become strongly anisotropic and develop nearly rectangular corners. The growth of $L(t)$ then saturates around $t/\tau\simeq10^4$, indicating arrested coarsening induced by the rotational bias.
}
    \label{Fig3}
\end{figure}
\subsection{Arrested coarsening dynamics}
To illustrate the dynamics generated by Eq.~\eqref{dm0final}, Fig.~\ref{Fig3} shows the time evolution for a $60\times60$ lattice with periodic boundary conditions, starting from random initial conditions. The upper three rows show snapshots of the magnetization for three different values of the rotational bias $\delta$ (see Eq.~\eqref{defdelta}), while the lower panel shows the corresponding characteristic length scale $L(t)$, whose definition is given in Appendix~\ref{app:lengthscale}. In all cases, the system initially undergoes phase separation and domain coarsening as shown in the first column.

For the unbiased case, $\delta=0$, the system continues to coarsen toward a macroscopically phase-separated stripe state. The late-time growth of the characteristic length is consistent with the scaling $L(t)\sim t^{1/4}$. This exponent is expected for conserved dynamics with a concentration dependent mobility which vanishes in the bulk phases \cite{DAI201685,PhysRevE.56.758}. Indeed, the discrete MF mobilities given by Eq.~\eqref{mobilities} vanish in the ordered bulk phases where neighboring magnetizations approach $m_i\simeq\pm1$. Since bulk diffusion is then suppressed inside the ordered domains, coarsening is expected to be dominated by interfacial transport, giving the surface-diffusion-limited growth law $L(t)\sim t^{1/4}$ rather than the Lifshitz--Slyozov law $L(t)\sim t^{1/3}$ associated with bulk diffusion \cite{DAI201685,LIFSHITZ196135}.

By contrast, for fully biased dynamics, $\delta=\pm1$, the coarsening process arrests. The domains become strongly anisotropic and develop nearly rectangular corners, and the growth of $L(t)$ saturates around $t/\tau\simeq10^4$. The saturation of $L(t)$ indicates that the rotational bias introduces an intrinsic length scale that prevents complete macroscopic phase separation within the MF dynamics. The resulting rectangular morphologies are qualitatively consistent with the microscopic simulations shown in \cite{wang2026edgecurrentsshapecondensates}. However, within the present one-site MF dynamics, the interfaces do not develop a systematic tilt w.r.t.~the $(x,y)$ axes as reported in \cite{wang2026edgecurrentsshapecondensates}. The apparent discrepancy is most likely due to the MF closure: by replacing the probability of a $2\times2$ block configuration by independent one-site Bernoulli probabilities, the closure neglects the short-range interfacial correlations and specific $2\times2$ motifs that determine the preferred microscopic interface orientation.
\section{Continuum limit}\label{sec:cont}
Next, we take the continuum limit of Eq.~\eqref{dm0final} following the same coarse-graining procedure as in \cite{10.21468/SciPostPhys.20.1.005}. We consider the thermodynamic limit $N_x,N_y\to\infty$ at fixed system size $\{L_x,L_y\}=\{N_x\ell,N_y\ell\}={\rm constant}$, so that the lattice spacing tends to zero, $\ell\to0$. In this limit, the discrete magnetization field is replaced by a smooth continuum field $m(\boldsymbol{x},t)$, defined as a local average over a box of linear size large compared to $\ell$. Since the coarse-grained field is smooth, finite differences may be expanded in spatial derivatives \cite{Penrose1991}. For example, the sum over nearest neighbors becomes
\begin{equation*}
\sum_{\langle ij\rangle} m_j(t)
\simeq
4m(\boldsymbol{x},t)+\ell^2\nabla^2 m(\boldsymbol{x},t)+\mathcal O(\ell^4).
\end{equation*}
In the following we measure lengths in units of $\ell$ and absorb these factors into the spatial derivatives. It is important to note that this gradient expansion also removes the explicit square-lattice structure of the original MF equations. Therefore it is expected that the resulting equation does \emph{not} reproduce the strongly rectangular morphologies observed in the full set of coupled MF lattice equations given by Eq.~\eqref{dm0final}. 
Applying the gradient expansion to Eq.~\eqref{dm0final} and only keeping the leading order terms (i.e. making a long-wavelength approximation) yields the following result\footnote{For a detailed derivation of Eq.~\eqref{finalresultcontinuum} starting from Eq.~\eqref{dm0final} we refer to the Mathematica notebook in \cite{BlomThieleZenodo2026}.}:
\begin{equation}
    \tau\frac{\partial m(\boldsymbol{x},t)}{\partial t}
    =
    \nabla\cdot\left[
    \frac{1-m^2(\boldsymbol{x},t)}{2}
    \nabla
    \frac{\delta \mathcal{F}_{\rm MF}[m(\boldsymbol{x},t)]}{\delta m(\boldsymbol{x},t)}
    \right]
    +
    \frac{J_x{+}J_y}{2}\delta\,
    \nabla\cdot\left[
    |\nabla m(\boldsymbol{x},t)|^2
    \boldsymbol{\varepsilon}
    \nabla m(\boldsymbol{x},t)
    \right],
    \label{finalresultcontinuum}
\end{equation}
where $\boldsymbol{\varepsilon}$ is the rotation matrix given by 
\begin{equation}
\boldsymbol{\varepsilon}
\equiv 
\begin{pmatrix}
0 & 1\\
-1 & 0
\end{pmatrix}
\label{Rmatrix}.
\end{equation}
The free energy functional $\mathcal{F}_{\rm MF}[m(\boldsymbol{x},t)]$ in Eq.~\eqref{finalresultcontinuum} is the continuum limit of Eq.~\eqref{FMF} given by
\begin{align}
    \!\!\!\!\mathcal{F}_{\rm MF}[m(\boldsymbol{x},t)]
    &\equiv
    \frac{1}{2}\!\int \!\! d\boldsymbol{x}\,
    \!\!\left[
    \Phi(m(\boldsymbol{x},t))
    {-}2(J_x{+}J_y)m(\boldsymbol{x},t)^2
    {-}m(\boldsymbol{x},t)[J_x\partial^2_x {+}J_y\partial^2_y]m(\boldsymbol{x},t) 
    \right]
    \nonumber \\
    &\stackrel{\rm p.i.}{=}
    \! \frac{1}{2}\! \int \!\! d\boldsymbol{x}\,
    \!\!\left[
    \Phi(m(\boldsymbol{x},t))
    {-}2(J_x{+}J_y)m(\boldsymbol{x},t)^2
    {+}J_x[\partial_x m(\boldsymbol{x},t)]^2{+}J_y[\partial_y m(\boldsymbol{x},t)]^2
    \right],
    \label{Fmfcontinuum}
\end{align}
where from the first to the second line we applied partial integration to the gradient term while assuming Neumann boundary conditions.

\textit{Relaxational term:} The first term on the right-hand side of Eq.~\eqref{finalresultcontinuum} describes Cahn--Hilliard-type relaxational dynamics with a magnetization-dependent mobility. For $\delta=0$ (i.e., $p=1/2$), the continuum equation therefore describes conserved gradient dynamics driven by the MF free energy $\mathcal{F}_{\rm MF}$. In this case the MF free energy is a Lyapunov functional, since for Neumann boundary conditions one obtains
\begin{equation*}
    \tau\frac{{\rm d}\mathcal{F}_{\rm MF}}{{\rm d}t}
    =\tau
    \int d\boldsymbol{x}\,
    \frac{\delta \mathcal{F}_{\rm MF}}{\delta m(\boldsymbol{x},t)}
    \frac{\partial m(\boldsymbol{x},t)}{\partial t}
    \underset{\delta=0}{\overset{\rm p.i.}{=}}
    -
    \int d\boldsymbol{x}\,
    \frac{1-m^2(\boldsymbol{x},t)}{2}
    \left|
    \nabla
    \frac{\delta \mathcal{F}_{\rm MF}}{\delta m(\boldsymbol{x},t)}
    \right|^2
    \leq 0 ,
\end{equation*}
where for the second equality we inserted Eq.~\eqref{finalresultcontinuum} and applied partial integration. Thus, in the absence of the chiral term, the dynamics monotonically decreases the MF free energy while conserving the total magnetization. Since the uniform free energy has two degenerate minima when $J_x+J_y>1/2$, the passive dynamics relaxes toward phase-separated states.

\textit{Chiral term:} The second contribution in Eq.~\eqref{finalresultcontinuum} may be interpreted as a chiral active interfacial current. Since $\boldsymbol{\varepsilon}\nabla m$ is obtained by rotating $\nabla m$ by $\pi/2$, it is everywhere orthogonal to the local gradient and therefore tangent to the level sets of $m(\boldsymbol{x},t)$. As a consequence, this nonequilibrium current flows along interfaces, i.e.\ in those spatial regions where $|\nabla m|$ is nonzero, while it vanishes in the bulk phases where the field is approximately uniform. This is consistent with the chiral current proposed in \cite{wang2026edgecurrentsshapecondensates}, where the chiral dynamics generates active currents along domain boundaries. The main difference is that the MF derivation presented here yields a current whose amplitude is controlled by the local sharpness of the interface, through $|\nabla m|^2$, but has no explicit dependence on the local interface orientation. In \cite{wang2026edgecurrentsshapecondensates}, by contrast, the active current is assumed to feature an explicitly angle-dependent amplitude. This difference most likely originates from the MF approximation, which neglects higher-order correlations generated by the microscopic block rotations. 

The chiral term also admits a representation in terms of the two-dimensional Poisson bracket $\{f,g\}\equiv\partial_x f\,\partial_y g-\partial_y f\,\partial_x g$ as:
\begin{equation*}
    \nabla\cdot\left[
    |\nabla m|^2
    \boldsymbol{\varepsilon}
    \nabla m
    \right]
    =
    \{
    |\nabla m|^2,m
    \} .
\end{equation*}
Equivalently, since
$\nabla\cdot\left[\boldsymbol{\varepsilon}\nabla m\right]=0$,
the chiral term can be written as an advection term
\begin{equation*}
    \nabla\cdot\left[
    |\nabla m|^2
    \boldsymbol{\varepsilon}
    \nabla m
    \right]
    =
    -
    \left[
    \boldsymbol{\varepsilon}
    \nabla|\nabla m|^2
    \right]\cdot\nabla m ,
\end{equation*}
with a divergence-free velocity field
$\nabla\cdot\left[\boldsymbol{\varepsilon}\nabla|\nabla m|^2\right]=0$.
This form shows that the active term alone transports the values of $m$ without changing their distribution. In particular, it conserves
$\int d\boldsymbol{x}\,\chi(m)$ for any sufficiently smooth function
$\chi(m)$.\footnote{
    More explicitly, considering only the chiral contribution, one obtains
    \begin{equation*}
        \left.
        \tau\frac{\rm d}{{\rm d}t}
        \int d\boldsymbol{x}\,\chi(m)
        \right|_{\rm chiral}
        =
        -
        \frac{J_x+J_y}{2}\delta
        \int d\boldsymbol{x}\,
        \chi'(m)
        \left[
        \boldsymbol{\varepsilon}
        \nabla|\nabla m|^2
        \right]\cdot\nabla m .
    \end{equation*}
    Using $\chi'(m)\nabla m=\nabla\chi(m)$ and
    $\nabla\cdot\left[\boldsymbol{\varepsilon}\nabla|\nabla m|^2\right]=0$,
    this can be written as
    \begin{equation*}
        \left.
        \tau\frac{\rm d}{{\rm d}t}
        \int d\boldsymbol{x}\,\chi(m)
        \right|_{\rm chiral}
        =
        -
        \frac{J_x+J_y}{2}\delta
        \int d\boldsymbol{x}\,
        \nabla\cdot\left[
        \chi(m)
        \boldsymbol{\varepsilon}
        \nabla|\nabla m|^2
        \right]
        =
        0 ,
    \end{equation*}
    where we used Neumann boundary conditions.
}
Thus, the active term does not by itself change the distribution of magnetization values, but only rearranges these values in space.

The absence of an explicit angular dependence also implies that the chiral contribution does not modify any radially symmetric steady state. Indeed, for a radially symmetric profile $m(\boldsymbol{x})=m_0(r)$ one has $
\nabla m_0(r)=m'_0(r)\boldsymbol{\hat{r}}$ and therefore $
\boldsymbol{\varepsilon}\nabla m_0(r)
=-m_0'(r)
\boldsymbol{\hat{\theta}}$
where $m'_0(r)\equiv {\rm d}m_0(r)/{\rm d}r$ and $(\boldsymbol{\hat{r}},\boldsymbol{\hat{\theta}})$ are the unit vectors in the radial and angular direction, respectively. Hence the active current $
|\nabla m_0(r)|^2\boldsymbol{\varepsilon}\nabla m_0(r)=-(m'_0(r))^3\boldsymbol{\hat{\theta}}$
has no radial component and is independent of the polar angle. Therefore it is divergence free,
\begin{equation*}
\nabla\cdot\left[|\nabla m_0(r)|^2\boldsymbol{\varepsilon}\nabla m_0(r)\right]
=
\frac{1}{r}\partial_\theta(-(m'_0(r))^3)
=
0.
\end{equation*}
Thus, any radially symmetric steady-state solution for $\delta=0$ remains a valid steady-state solution also for $\delta\neq0$. However, the chiral current can affect the stability of such states with respect to non-radial perturbations, as we will see in the next section. Note that the active current also does not affect the spinodal instability of a spatially uniform state, as a perturbation of the form $\delta m \exp(i \mathbf{k}\cdot\boldsymbol{x}+\lambda  t)$ vanishes due to the antisymmetric rotation matrix $\boldsymbol{\varepsilon}$. The nonequilibrium current therefore affects the dynamics only beyond linear order, most prominently in interfacial regions, e.g., in the stability of interfaces against non-radial deformations.
\subsection{Chiral fingering instability}\label{sec:chiralinstability}
We next analyze the stability of radially symmetric steady states with respect to angular perturbations. For simplicity we restrict the analysis to isotropic coupling strengths $J_x=J_y=J$. Furthermore, we treat $\delta$ as an effective continuum driving parameter and allow values outside the microscopic range $\delta\in[-1,1]$. This is justified because the strength of the chiral current is controlled by the product $J\delta$. Large effective driving could equivalently be obtained by increasing $J$, but large $J$ drives the coexisting bulk phases close to $m=\pm1$, where the logarithmic terms in the mean-field free energy cause numerical stiffness. We therefore keep $J$ relatively small and access large effective chiral driving by increasing $\delta$. 

As discussed above, any radially symmetric steady state of the equilibrium dynamics for fixed $J$ and $\delta=0$ remains a steady state in the presence of the chiral current with $\delta \neq 0$. We therefore first numerically calculate radially symmetric steady profiles $m_0(r)$ which exist for $J>1/4$, where the homogeneous state with $m=0$ becomes unstable and two locally coexisting bulk phases appear. The radial profile $m_0(r)$ is obtained by solving the stationary equation at fixed conserved mean magnetization which is further explained in Appendix \ref{app:angularstability}.

To determine whether this radially symmetric state is stable against shape deformations, we perturb it by angular Fourier modes,
\begin{equation}
m(r,\theta,t)
=
m_0(r)
+\eta\,
u_n(r)
\exp\left(\sigma_n t+i n\theta\right),
\label{eq:angularperturbation}
\end{equation}
where $|\eta|\ll1$, $n\in\mathbb{N}$ is the angular mode number, $u_n(r)$ is the radial eigenfunction, and $\sigma_n$ is the corresponding eigenvalue. The angular mode number $n$ has a direct geometric interpretation: it determines how many periods the perturbation has around the circular interface. Thus, an unstable mode with angular number $n$ produces an interfacial deformation with $n$ protrusions, or fingers. The real part of $\sigma_n$ determines the growth or decay of this deformation, while the imaginary part describes a drift of the angular phase and therefore corresponds to a rotation of the interfacial pattern around the circular domain.\footnote{Strictly speaking, a real perturbation can be constructed as a superposition of complex conjugate angular modes. Depending on the relative amplitudes and phases, this may appear as an oscillating standing deformation rather than a purely rotating one. However, because the chiral contribution breaks reflection symmetry, the generic unstable eigenmode has a definite angular drift and is naturally interpreted as a rotating interfacial deformation.} The $n=0$ mode corresponds to a radially symmetric growth or shrinkage perturbation, while $n=1$ corresponds to a translation of the radially symmetric state. Shape instabilities are therefore associated with modes $n\geq 2$. Substituting Eq.~\eqref{eq:angularperturbation} into Eq.~\eqref{finalresultcontinuum} and retaining only terms linear in $u_n$ yields a radial eigenvalue problem of the form
\begin{equation*}
\tau \sigma_n u_n
=
\hat{\mathcal{L}}_n u_n ,
\end{equation*}
where $\hat{\mathcal{L}}_n$ is a non-Hermitian radial differential operator defined in Eq.~\eqref{app:eq:eigenvalueproblem}. For each angular mode $n$, we solve the resulting one-dimensional eigenvalue problem numerically. A radially symmetric state is linearly unstable to shape perturbations if $\max_{n\geq 2}\operatorname{Re}\sigma_n > 0$. The angular mode for which $\operatorname{Re}\sigma_n$ is maximal gives the dominant unstable deformation of the interface and predicts the number of fingers that grow fastest during the early stage of the instability.
\begin{figure}[t!]
    \centering
    \includegraphics[width=\linewidth]{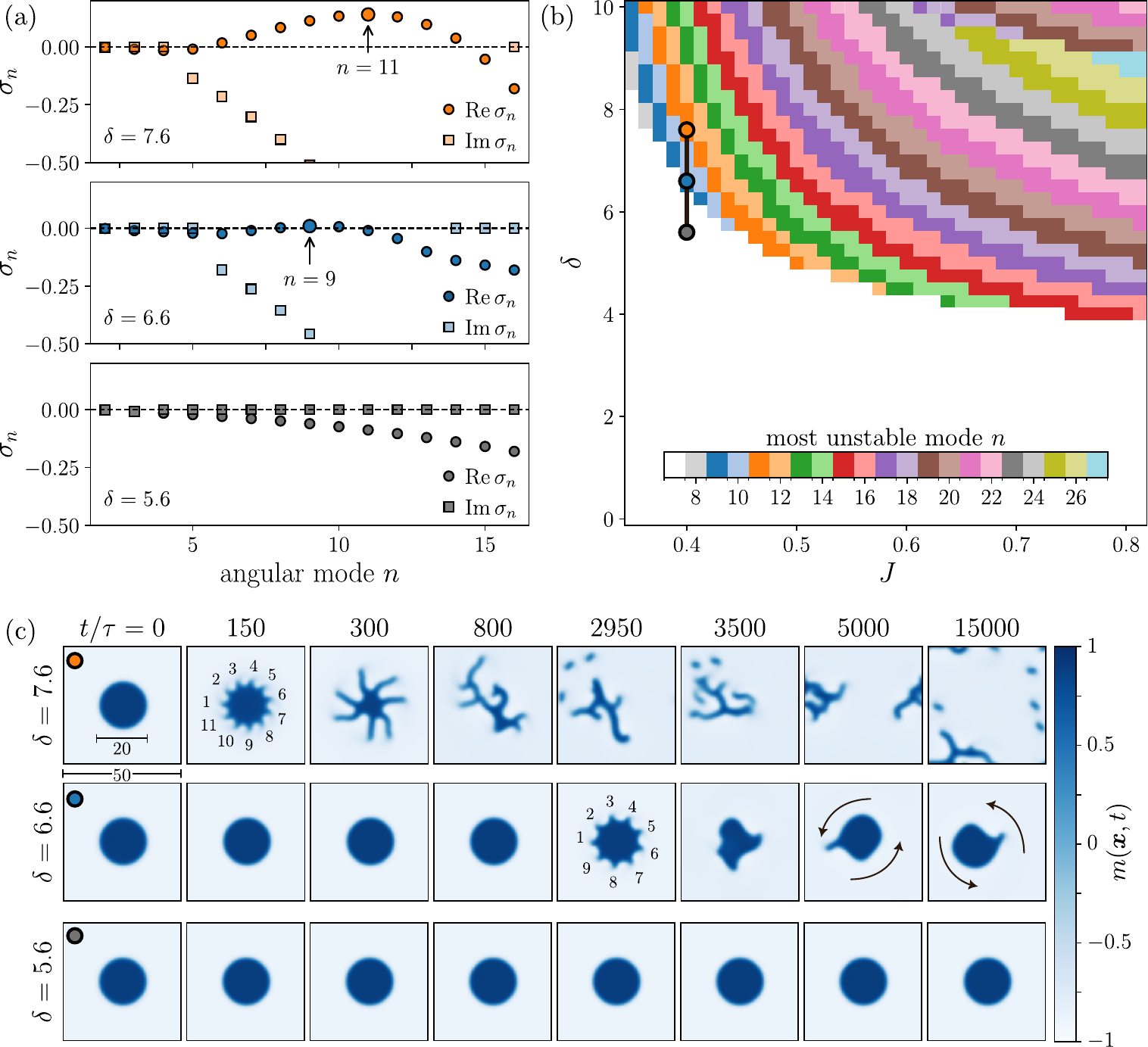}
    \caption{\textbf{Chiral fingering instability of radially symmetric states.}
(a) Dispersion relations for angular perturbations of a radially symmetric state at fixed $J=0.4$ and, from top to bottom, $\delta=\{7.6,6.6,5.6\}$. Calculations are performed in a square domain of side length $L=50$, with an initially circular domain of radius $R_0=10$. For the two larger values of $\delta$, an intermediate band of angular modes has positive growth rate ${\rm Re} \, \sigma_n$ and nonzero frequency ${\rm Im} \, \sigma_n$, indicating an oscillatory interfacial instability. The fastest-growing modes are $n=11$ for $\delta=7.6$ and $n=9$ for $\delta=6.6$. For $\delta=5.6$, all angular modes are stable. (b) Stability diagram in the $(J,\delta)$ plane. The color indicates the angular mode $n$ with the largest growth rate. In the white region all growth rates are negative and radially symmetric states are stable against angular perturbations, whereas in the colored region radially symmetric states are unstable and complex. The marked points correspond to the three dispersion relations shown in panel (a). (c) Time simulations for the same three parameter values as in panel (a). For $\delta=7.6$ and $\delta=6.6$, the initially radially symmetric state develops interfacial fingers whose number matches the fastest-growing mode predicted by the linear stability analysis, namely $n=11$ and $n=9$, respectively. At later times these protrusions coarsen and grow further, and the circular domain either breaks up into an irregular morphology ($\delta{=}7.6$) or evolves into a domain with a single arm that rotates counterclockwise ($\delta{=}6.6$). For $\delta{=}5.6$, the radially symmetric state remains stable. In all panels $\tau=1$. Animations of the time simulations can be found in \cite{BlomThieleZenodo2026}.
}
    \label{Fig4}
\end{figure}
\begin{figure}[t!]
    \centering
    \includegraphics[width=\linewidth]{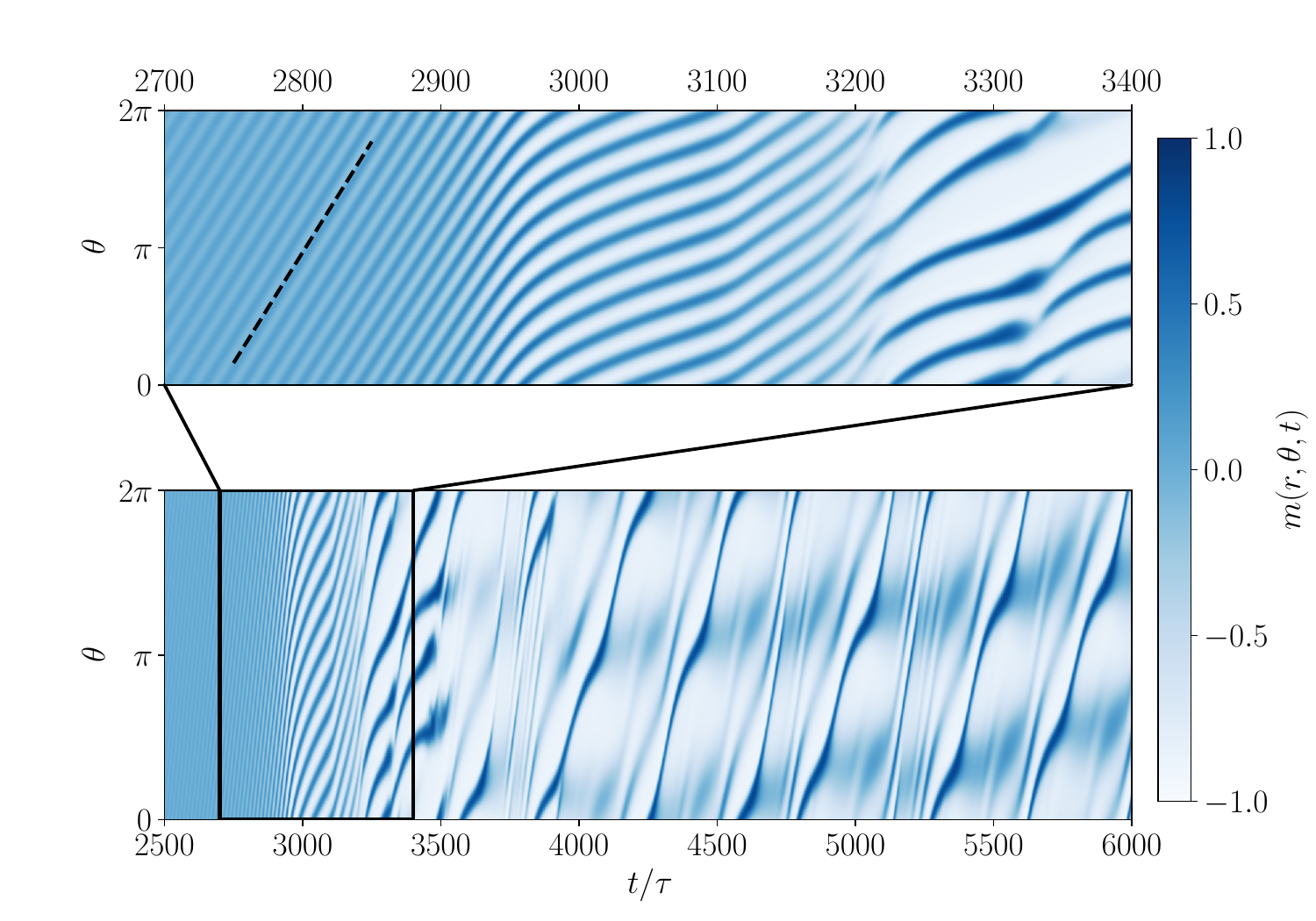}
    \caption{\textbf{Spacetime representation of the chiral fingering instability and radially asymmetric rotating state for $\delta=6.6$ and $J=0.4$.} The field $m(r,\theta,t)$ is sampled at fixed radius $r=10$ around the instantaneous center of the domain and plotted as a function of time and angle $\theta$. The lower panel shows the full time interval, while the upper panel shows a zoom into the time window from $t/\tau=2700$ to $t/\tau=3400$. The dashed black line in the zoomed panel indicates the angular drift expected from the imaginary part of the linear growth rate of the dominant $n=9$ mode. Using $\operatorname{Im}\sigma_9=-0.45635$ (see Fig.~\ref{Fig4}a) gives the phase velocity $\dot{\theta}=-\operatorname{Im}\sigma_9/9\simeq 0.0507$, in agreement with the slope of the spacetime pattern for short times. After the initial stage of the fingering instability, the spacetime plot shows that the phase velocity first slow down, then speed up again, and finally the fingers coarsen into a single rotating arm.
 }
    \label{Fig5}
\end{figure}

Figure~\ref{Fig4} summarizes the resulting linear stability analysis. For fixed $J=0.4$, Fig.~\ref{Fig4}a shows that sufficiently strong chiral driving destabilizes an intermediate band of angular modes. The fastest-growing modes are $n=11$ for $\delta=7.6$ and $n=9$ for $\delta=6.6$, predicting the initial formation of $11$ and $9$ interfacial fingers, respectively. Since the corresponding eigenvalues are complex, the growing deformation also rotates around the circular domain. We therefore refer to this finite-wavelength oscillatory instability as a chiral fingering instability. For $\delta=5.6$, all angular modes have negative real part and the state is linearly stable. Related fingering instabilities have been discussed for active nematic droplets, where active stresses and interfacial anchoring destabilize the droplet boundary above a critical activity \cite{Alert_2022}. Here, by contrast, the instability occurs in a conserved scalar theory and is driven by a chiral interfacial current.

Repeating the calculation throughout the $(J,\delta)$ plane gives the stability diagram in Fig.~\ref{Fig4}b. The white region denotes parameters for which all angular perturbations decay, while the colored region indicates that at least one angular mode has positive real part. The color gives the mode with the largest growth rate and therefore predicts the number of interfacial fingers that initially form. Moving along the instability threshold the number of fingers $n$ monotonically increases with increasing $J$. Direct simulations of the continuum equation, shown in Fig.~\ref{Fig4}c, confirm this prediction: for $\delta=7.6$ and $\delta=6.6$, the initially circular domain develops approximately $11$ and $9$ protrusions, respectively, while for $\delta=5.6$ it remains stable.

To examine the rotational dynamics of the unstable fingers more directly, Fig.~\ref{Fig5} shows a spacetime plot of $m(r,\theta,t)$ for $\delta=6.6$, sampled at a fixed radius around the instantaneous center of the domain. At early times, the slope of the spacetime pattern agrees with the phase velocity $\dot{\theta}=-\operatorname{Im}\sigma_9/9$ predicted by the complex eigenvalue of the dominant $n=9$ mode (see black dashed line). This confirms that the initial rotation of the fingers is governed by the linear chiral fingering instability. At later times, however, the deformation is no longer small, and the evolution is no longer described by the linear eigenmode alone. As also shown in Fig.~\ref{Fig5}, the fingers first slow down, then speed up again, and finally coarsen into a single radially asymmetric rotating arm for $\delta=6.6$. For the larger value $\delta=7.6$, by contrast, the protrusions continue to grow and interact, eventually destroying the regular circular shape and producing an irregular morphology. Hence, the nonlinear evolution of the chiral fingering instability can lead either to coherent time-periodic dynamics or to an irregular morphology, depending on the strength of the chiral drive.
\section{Conclusion}
In this work, we have derived a MF continuum theory for the chiral Ising model introduced in \cite{wang2026edgecurrentsshapecondensates}. Starting from the microscopic master equation for clockwise and counterclockwise rotations of $2\times2$ spin blocks, we have obtained closed MF lattice equations for the spatially resolved magnetization (see Eq.~\eqref{dm0final}). These equations conserve the total magnetization and explicitly show how a microscopic rotational bias enters the coarse-grained dynamics.

On the lattice level, we found that biased block rotations qualitatively modify coarsening dynamics (see Fig.~\ref{Fig3}). In contrast to the unbiased dynamics, which continues to coarsen, chiral driving produces anisotropic domain growth, nearly rectangular morphologies, and arrested coarsening. Taking the continuum limit, we obtained a conserved active field theory consisting of a passive Cahn-Hilliard contribution and a chiral interfacial current (see Eq.~\eqref{finalresultcontinuum}). This current is tangent to level sets of the magnetization and is localized at interfaces. Compared with the phenomenological current discussed in \cite{wang2026edgecurrentsshapecondensates}, the MF current has no explicit dependence on the local interface orientation, suggesting that such orientational information is encoded in correlations neglected by the one-site mean-field approximation.

Using the continuum theory, we analyzed the angular stability of radially symmetric phase-separated states (see Figs.~\ref{Fig4} and \ref{Fig5}). Although the chiral current does not modify these steady states themselves, it can destabilize them with respect to non-radial perturbations. The resulting instability has a finite angular wavelength and complex eigenvalues, so that interfacial fingers grow while rotating around the domain. Direct simulations confirm that the fastest-growing angular mode predicts the number of fingers formed at early times. We therefore identify this mechanism as a chiral fingering instability.

Several questions remain for future work. First, the transition from radially symmetric states to rotating radially asymmetric states or irregular morphologies appears to be subcritical, and the resulting irregular states may be related to chaotic dynamics. This should be tested more systematically, for example by analyzing long-time statistics and Lyapunov spectra. Second, numerical continuation could be used to track stable and unstable solution branches and clarify how the subcritical transition emerges. Finally, the dependence of the instability on curvature suggests a possible mechanism for chirality-induced domain-size selection: if smaller circular domains become unstable more easily, chiral driving could select a characteristic domain size by destabilizing domains outside a stable size range.
\appendix
\section{Nonnegativity of the mobilities}\label{AppendixMobilities}
 The mobilities $\mathcal{M}^{-,-}_{0,{\rm d}}$, $\mathcal{M}^{-,-}_{0,x}$, and $\mathcal{M}^{-,-}_{0,y}$ reported in Eq.~\eqref{mobilities} are manifestly nonnegative, since for all $\alpha,\beta\in[-1,1]$ and $\gamma\in\mathbb{R}$ we have $1-\alpha \beta\geq 0$ and $1-\alpha \beta+(\alpha-\beta)\tanh{(\gamma)}\geq 0$. For the mobility $\mathcal{M}^{-,-}_{0,{\rm a}}$ we note that it depends linearly on the quantity $
T \equiv \tanh(\langle h_0 \rangle + \langle h_8 \rangle - \langle h_1 \rangle - \langle h_7 \rangle)$. For a linear function in $T$, the minimum and maximum occur at the extreme values, which are $T = \pm 1$. Evaluating $\mathcal{M}^{-,-}_{0,{\rm a}}$ at these points, we find
\begin{equation*}
\mathcal{M}^{-,-}_{0,{\rm a}}\big|_{T=\pm1} = (1\mp m_0)(1\pm m_1)(1\pm m_7)(1\mp m_8) \ge 0.
\end{equation*}
Since $\mathcal{M}^{-,-}_{0,{\rm a}}$ is linear in $T$, it cannot take smaller values in between $T=-1$ and $T=1$, and therefore it follows that $\mathcal{M}^{-,-}_{0,{\rm a}}\geq 0$.
\section{Characteristic length scale}\label{app:lengthscale}
In order to quantify the coarsening dynamics of Eq.~\eqref{dm0final}, we define a characteristic length scale from the equal-time structure factor of the magnetization. For a lattice configuration $m_i(t)$, we first subtract the spatial average,
\begin{equation*}
    \phi_i(t)
    =
    m_i(t)-\bar m,
    \qquad
    \bar m= 
    \frac{1}{N}\sum_i m_i(t),
\end{equation*}
where $N=N_xN_y$ is the total number of lattice sites. Since the dynamics conserves the total magnetization, $\bar m$ is constant in time. We then compute the discrete Fourier transform
\begin{equation*}
    \hat{\phi}_{\mathbf q}(t)
    =
    \sum_i
    \phi_i(t)
    \exp\left(-i\mathbf q\cdot\mathbf r_i\right),
\end{equation*}
where due to the periodic boundary conditions the allowed wavevectors are $\mathbf q=2\pi (n_x/N_x,n_y/N_y)^{\rm T}$ with $n_x$ and $n_y$ positive integers. The corresponding structure factor is defined as
\begin{equation*}
    S(\mathbf q,t)
    =
    |\hat{\phi}_{\mathbf q}(t)|^2 .
\end{equation*}
The characteristic wavenumber is obtained from the first moment of the structure factor,
\begin{equation*}
    q_{\rm av}(t)
    \equiv
    \frac{
    \sum_{\mathbf q\neq 0}
    |\mathbf q| S(\mathbf q,t)
    }{
    \sum_{\mathbf q\neq 0}
    S(\mathbf q,t)
    },
\end{equation*}
where the zero mode is excluded because it contains only the conserved spatial average. We define the characteristic length scale as
\begin{equation}
    L(t)
    \equiv 
    \frac{2\pi}{q_{\rm av}(t)}
    .
    \label{LengthScaleDefinition}
\end{equation}
With this convention, $L(t)$ has the interpretation of a typical wavelength associated with the dominant domain pattern. In particular, for a pattern dominated by a single Fourier mode with wavenumber $q_\ast$, Eq.~\eqref{LengthScaleDefinition} gives $L(t)\simeq 2\pi/q_\ast$.
\section{Angular linear stability analysis }
\label{app:angularstability}
In this appendix we derive the radial eigenvalue problem used in the angular linear stability analysis of Sec.~\ref{sec:chiralinstability}. We restrict ourselves to the isotropic case $J_x=J_y=J$. As shown in the main text, the chiral current does not modify radially symmetric steady states. Hence the stationary profile $m_0(r)$ is determined by the passive part of Eq.~\eqref{finalresultcontinuum}. Transforming to polar coordinates the steady state satisfies
\begin{equation}
\frac{1}{r}
\frac{{\rm d}}{{\rm d}r}
\left[
r M_0
\frac{{\rm d}\mu_0}{{\rm d}r}
\right]=0,
\label{app:eq:m0stationary}
\end{equation}
where the chemical potential of the stationary profile $m_0(r)$ reads
\begin{equation*}
\mu_0(r)
=
\arctanh(m_0)
-4Jm_0
-J
\left[
m_0''
+
\frac{m_0'}{r}
\right],
\end{equation*}
and $M_0\equiv M(m_0(r))$ with the mobility given by
\begin{equation*}
M(m)\equiv (1-m^2)/2.
\end{equation*}
For no-flux boundary conditions, Eq.~\eqref{app:eq:m0stationary} implies that $\mu_0$ is spatially constant. Thus $m_0(r)$ solves the equation $\mu_0(r)=\mu_\ast$, where the constant $\mu_\ast$ is fixed by the conserved total mass. The radial profile is supplemented by the natural boundary conditions $m'_0(0)=m'_0(R)=0$. We now perturb this radially symmetric state by angular Fourier modes
\begin{equation*}
m(r,\theta,t)
=
m_0(r)
+\eta\,
u_n(r)
\exp\left(\sigma_n t+i n\theta\right),
\end{equation*}
For later use, we introduce the radial Laplacian acting on the $n$th angular mode,
\begin{equation*}
\Delta_n u
\equiv
u''
+
\frac{u'}{r}
-
\frac{n^2}{r^2}u.
\end{equation*}
The corresponding linear perturbation of the chemical potential reads $\mu=\mu_0+\eta \, \mu_n(r)\exp(\sigma_n t+in\theta)+\mathcal{O}(\eta^2)$, where
\begin{equation*}
\mu_n(r)
=
\left[
\frac{1}{1-m^2_0(r)}
-
4J
-
J\Delta_n
\right] u_n(r).
\end{equation*}
Since the unperturbed state has constant chemical potential, the linearization of the passive Cahn--Hilliard part gives
\begin{equation*}
\nabla\cdot
\left[
M(m)\nabla \mu
\right]
=
\eta \exp(\sigma_n t+in\theta) \hat{\mathcal{D}}_n\mu_n(r)
,
\end{equation*}
where we have introduced the operator
\begin{equation*}
\hat{\mathcal{D}}_n
\equiv
M_0\Delta_n+M_0'\frac{\rm d}{{\rm d}r}.
\end{equation*}
The perturbation of the mobility does not contribute at linear order because it is multiplied by $\nabla\mu_0=0$. It remains to linearize the chiral contribution. Using $\nabla = \boldsymbol{\hat{r}}\partial_r+\boldsymbol{\hat{\theta}}(1/r)\partial_\theta$ and therefore $|\nabla m_0|^2=[m_0'(r)]^2$, one obtains
\begin{equation*}
\nabla\cdot
\left[
|\nabla m|^2
\boldsymbol{\varepsilon}\nabla m
\right]
=
2\eta \exp(\sigma_n t+in\theta)
\frac{in}{r}
\left[
m_0'(r)m_0''(r)u_n(r)
-
[m_0'(r)]^2u_n'(r)
\right].
\end{equation*}
Equivalently, this can be written in the compact form
\begin{equation*}
\nabla\cdot
\left[
|\nabla m|^2
\boldsymbol{\varepsilon}\nabla m
\right]
=
-2\eta \exp(\sigma_n t+in\theta)
\frac{in}{r}
[m_0'(r)]^3
\frac{{\rm d}}{{\rm d}r}
\left[
\frac{u_n(r)}{m_0'(r)}
\right].
\end{equation*}
This form makes clear that the chiral contribution vanishes for perturbations proportional to $m_0'(r)$, corresponding to a pure normal displacement of the interface, which is forbidden by the conservation law. Combining the passive and chiral contributions, we obtain from Eq.~\eqref{finalresultcontinuum} the radial eigenvalue problem
\begin{equation}
\tau\sigma_n u_n
=
\hat{\mathcal{D}}_n
\mu_n(r)
-
2J\delta
\frac{in}{r}
[m_0'(r)]^3
\frac{{\rm d}}{{\rm d}r}
\left[
\frac{u_n(r)}{m_0'(r)}
\right]\equiv \hat{\mathcal{L}}_nu_n.
\label{app:eq:eigenvalueproblem}
\end{equation}
The term proportional to $J\delta$ is imaginary and makes the operator non-Hermitian. Therefore, the eigenvalues are in general complex as shown in Fig.~\ref{Fig4}a.

To obtain Fig.~\ref{Fig4}a,b we solved Eq.~\eqref{app:eq:eigenvalueproblem} numerically using a spatial discretization of the radial eigenvalue problem. For each angular mode $n$, this yields a matrix eigenvalue problem which we solve numerically using Scipy \cite{2020SciPy-NMeth}. The leading eigenvalue for each $n$ determines the growth rate $\operatorname{Re}\sigma_n$ and frequency $\operatorname{Im}\sigma_n$ of the corresponding angular perturbation. The most unstable mode is obtained by comparing these leading growth rates over the shape modes $n\geq2$.
\section{Time simulations}
We performed time simulations of Eq.~\eqref{finalresultcontinuum} on a periodic square domain of side length $L=50$ using a pseudo-spectral method. The field $m(\boldsymbol{x},t)$ was discretized on an $N\times N$ grid with $N=200$, and spatial derivatives were computed in Fourier space. Throughout the simulations we used $\tau=1$. The time integration was performed with a semi-implicit Euler scheme. In Fourier space, the right-hand side was split into a linear contribution and a nonlinear remainder,
\begin{equation*}
    \partial_t \hat{m}_{\boldsymbol{k}}
    =
    \Lambda_{\boldsymbol{k}}
    \hat{m}_{\boldsymbol{k}}
    +
    \hat{\mathcal{N}}_{\boldsymbol{k}}(t) .
\end{equation*}
Here $\Lambda_{\boldsymbol{k}}$ denotes the linear part given by
\begin{equation*}
    \Lambda_{\boldsymbol{k}}
    \equiv
    -\frac{1}{2}(1-4J)|\boldsymbol{k}|^2
    -
    \frac{1}{2}J|\boldsymbol{k}|^4,
\end{equation*}
while $\hat{\mathcal{N}}_{\boldsymbol{k}}$ is the Fourier transform of the remaining nonlinear terms, which reads
\begin{equation*}
    \mathcal{N}[m]
    \equiv
    J\nabla\cdot
    \left[
        2m^2\nabla m
        +
        m^2\nabla\nabla^2 m/2
        +
        \delta |\nabla m|^2\boldsymbol{\varepsilon}\nabla m
    \right].
\end{equation*}
The linear part was treated implicitly, whereas the nonlinear part was treated explicitly. This gives the update
\begin{equation*}
    \hat{m}_{\boldsymbol{k}}(t+\Delta t)
    =
    \frac{
    \hat{m}_{\boldsymbol{k}}(t)
    +
    \Delta t\,
    \hat{\mathcal{N}}_{\boldsymbol{k}}(t)
    }{
    1-\Delta t\,\Lambda_{\boldsymbol{k}}
    },
\end{equation*}
with time step $\Delta t=5\times 10^{-3}$. The implicit treatment of the linear part improves numerical stability, because this part contains the stiff high-wavenumber $|\boldsymbol{k}|^4$ contribution coming from the fourth-order derivative term \cite{doi:10.1137/1.9781611978209}.  

For the simulations shown in Figs.~\ref{Fig4}c and \ref{Fig5}, we initialized the system with a radially symmetric domain of radius $R_0=10$. The bulk values inside and outside the domain were chosen using the bulk free-energy density $
f_{\rm b}(m)
\equiv 
\Phi(m)/2
-
2Jm^2$ with $\Phi(m)$ given by Eq.~\eqref{MFentropy}.
The corresponding bulk chemical potential and pressure are
\begin{equation*}
\mu_{\rm b}(m)
\equiv 
\frac{{\rm d} f_{\rm b}}{{\rm d}m}
=
\operatorname{arctanh}(m)-4Jm,
\qquad
p(m)
\equiv 
m\mu_{\rm b}(m)-f_{\rm b}(m).
\end{equation*}
The values $m_{\rm in}$ and $m_{\rm out}$ were then chosen such that the chemical potentials are equal,
\begin{equation*}
\mu_{\rm b}(m_{\rm in})
=
\mu_{\rm b}(m_{\rm out}),
\end{equation*}
and such that the pressure difference equals the Laplace pressure,
\begin{equation*}
p(m_{\rm in})-p(m_{\rm out})
=
\frac{\gamma}{R_0}.
\end{equation*}
Here $\gamma$ is the planar line tension corresponding to the same value of $J$. It is computed from the coexisting bulk phases $m_-$ and $m_+$ of a flat interface as \cite{10.1063/1.1744102,PhysRevResearch.5.013135}
\begin{equation*}
\gamma
=
\int_{m_-}^{m_+}
dm 
\sqrt{
2J
\left[
f_{\rm b}(m)-f_{\rm b}(m_+)
\right]
} .
\end{equation*}
Here, $m_-$ and $m_+$ denote the two planar coexistence values, satisfying $\mu_b(m_-)=\mu_b(m_+)$ and $p(m_-)=p(m_+)$. The initial interface was smoothed over a finite width using a hyperbolic tangent profile. To test the linear stability of the radially symmetric state in direct simulations, we perturbed the initial radius according to $R(\theta)=R_0[1+\eta\cos(n\theta)]$ with $\eta=10^{-7}$ with $n$ determined by the fastest growing angular mode.
\section*{Acknowledgements}
We thank Daniel Greve and Florian Voss for fruitful discussions. 
\section*{Data availability}
The data and python codes used in this study as well as a supplemental video accompanying Fig.~\ref{Fig4} are publicly available on \cite{BlomThieleZenodo2026}.
%
\end{document}